\documentclass{article}
\usepackage{arxiv}

\usepackage{hyperref}       %
\usepackage{soul}
\usepackage{subcaption}
\usepackage{url}
\usepackage[utf8]{inputenc}
\usepackage{caption}
\usepackage{graphicx}
\usepackage{amsmath}

\usepackage{amssymb}
\usepackage{amsthm}
\usepackage{booktabs}
\usepackage{xcolor}
\usepackage{algorithmicx}
\usepackage{siunitx}
\usepackage[linesnumbered,ruled,vlined]{algorithm2e}
\usepackage{pgfplots}
\usepackage{pgfplotstable}
\pgfplotsset{compat=1.18}
\usepgfplotslibrary{statistics}

\usepackage{tabularx}
\usepackage[switch]{lineno}
\usepackage{comment}

\usepackage{bm}
\usepackage{amssymb}
\usepackage{pifont}
\usepackage[table]{xcolor}
\newtheorem{theorem}{Theorem}

\title{CONSENT: A Negotiation Framework for Leveraging User Flexibility in Vehicle-to-Building Charging under Uncertainty}

\author{
  {Rishav Sen}\\
  Vanderbilt University\\
  Nashville, USA\\
  \texttt{rishav.sen@vanderbilt.edu}
  \And
  {Fangqi Liu}\\
  Vanderbilt University\\
  Nashville, USA\\
  \texttt{fangqi.liu@vanderbilt.edu}
  \And
  {Jose Paolo Talusan}\\
  Vanderbilt University\\
  Nashville, USA\\
  \texttt{jose.paolo.talusan@vanderbilt.edu}
  \And
  {Ava Pettet}\\
  Nissan Advanced Technology Center -- Silicon Valley\\
  Santa Clara, USA\\
  \texttt{ava.pettet@nissan-usa.com}
  \And
  {Yoshinori Suzue}\\
  Nissan Advanced Technology Center -- Silicon Valley\\
  Santa Clara, USA\\
  \texttt{yoshinori.suzue@nissan-usa.com}
  \And
  {Mark Bailey}\\
  Nissan Advanced Technology Center -- Silicon Valley\\
  Santa Clara, USA\\
  \texttt{mark.bailey@nissan-usa.com}
  \And
  {Ayan Mukhopadhyay}\\
  William \& Mary\\
  Williamsburg, USA\\
  \texttt{amukhopadhyay@wm.edu}
  \And
  {Abhishek Dubey}\\
  Vanderbilt University\\
  Nashville, USA\\
  \texttt{abhishek.dubey@vanderbilt.edu}
}
\date{}

\renewcommand{\shorttitle}{CONSENT: A Negotiation Framework for V2B Charging}

\hypersetup{
pdftitle={CONSENT: A Negotiation Framework for Leveraging User Flexibility in Vehicle-to-Building Charging under Uncertainty},
pdfsubject={},
pdfauthor={},
pdfkeywords={},
}
\begin{document}

\maketitle 

\begin{abstract}
The growth of Electric Vehicles (EVs) creates a conflict in vehicle-to-building (V2B) settings between building operators, who face high energy costs from uncoordinated charging, and drivers, who prioritize convenience and a full charge. To resolve this, we propose a negotiation-based framework that, by design, guarantees voluntary participation, strategy-proofness, and budget feasibility. It transforms EV charging into a strategic resource by offering drivers a range of incentive-backed options for modest flexibility in their departure time or requested state of charge (SoC).
Our framework is calibrated with user survey data and validated using real operational data from a commercial building and an EV manufacturer. Simulations show that our negotiation protocol creates a mutually beneficial outcome: lowering the building operator's costs by over 3.5\% compared to an optimized, non-negotiating smart charging policy, while simultaneously reducing user charging expenses by 22\% below the utility's retail energy rate.
By aligning operator and EV user objectives, our framework provides a strategic bridge between energy and mobility systems, transforming EV charging from a source of operational friction into a platform for collaboration and shared savings.
\end{abstract}


\section{Introduction}
\label{sec:introduction}

\begin{figure*}[htp]
    \centering
    \includegraphics[width=0.9\linewidth]{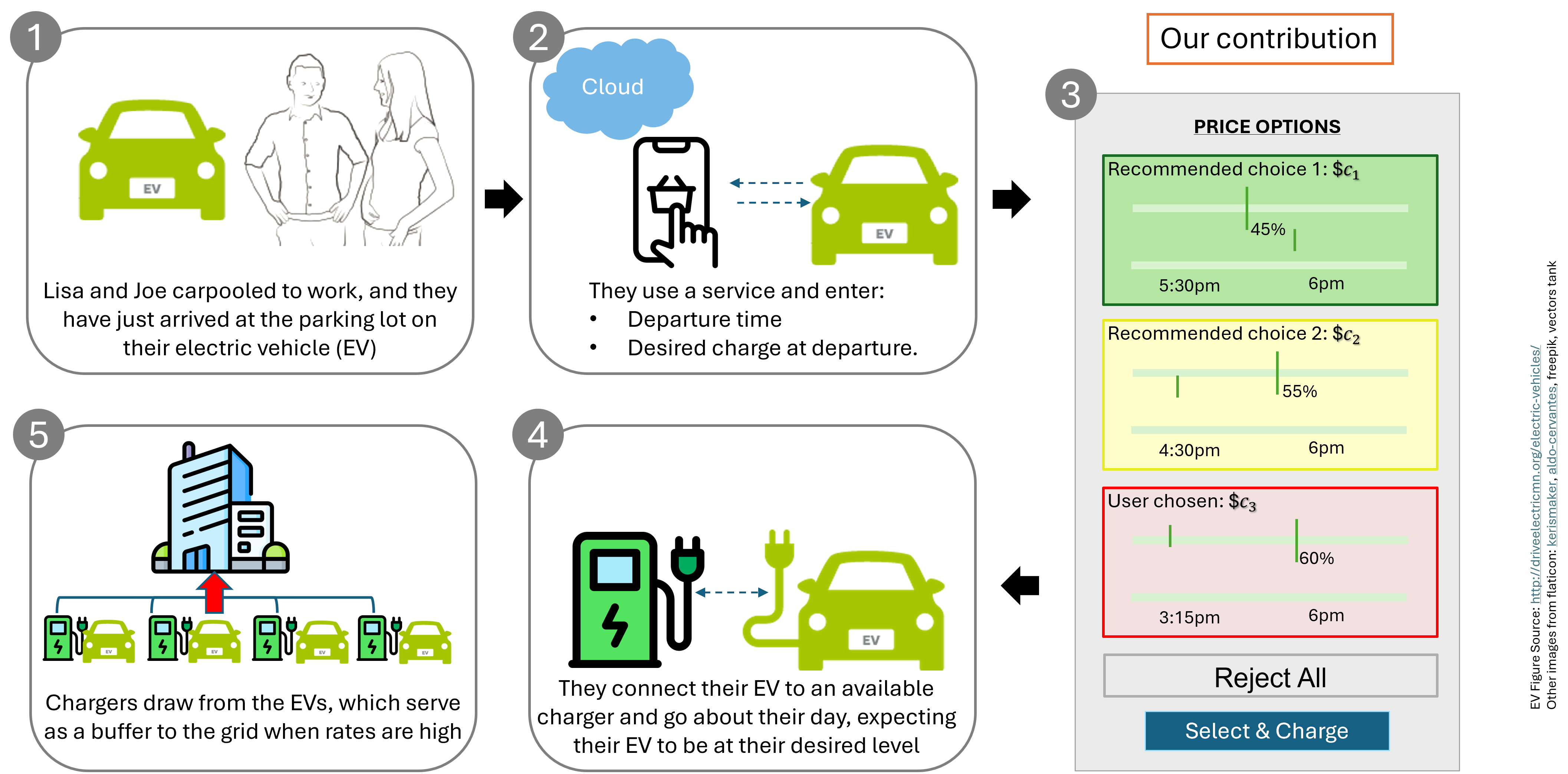}
    \caption{The CONSENT framework structures EV charging as a simple five-step interaction between the user and the building's energy management system. Users arrive, submit their baseline request, and the system's key contribution: \textbf{Incentive Generation} creates price–flexibility options based on forecasted load and energy prices (Panel~3). These options offer monetary discounts in exchange for modest flexibility in SoC or departure time. The user selects one option, after which the MPC engine executes the negotiated plan while ensuring the vehicle meets the agreed SoC and supporting building-level peak shaving.}
    \label{fig:intro}
    \vspace{-0.4cm}
\end{figure*}

The rapid growth of electric vehicles (EVs), fueled by sustainability goals and advances in battery technology~\cite{al2019review, wu2023electric}, presents a significant challenge for energy management. When EV charging is uncoordinated, it can trigger sharp demand spikes, inflate energy costs, and threaten grid stability~\cite{liu2021ev}. This has spurred research into coordinated charging, particularly in the vehicle-to-building (V2B) context where operators can leverage parked EVs for peak shaving~\cite{wang2018ev}. However, creating an effective V2B system requires navigating a set of deeply interconnected challenges that span both technical control and human behavior.

At the core of the control problem lies the interplay between {uncertainty} and {long-term rewards}~\cite{liu2025}. A charging controller must operate with incomplete knowledge of future building loads and EV arrivals, yet its decisions are judged against long-term objectives like minimizing the monthly peak demand charge, a sparse and delayed reward. This makes myopic, reactive control insufficient. This technical challenge is further complicated by the need for {incentive alignment}~\cite{bae2021inducing}. The operator’s goal of deferring charging to minimize electricity costs, driven by both energy prices and monthly peak-power demand charges, directly conflicts with the user's desire for convenience and a full battery~\cite{he2022optimal}. A negotiation must therefore quantify the value of a user’s flexibility, but this value is itself a function of an uncertain future. Finally, any negotiation is vulnerable to {strategic misreporting}, where users may exaggerate their needs to extract larger incentives, undermining the system's integrity. Addressing these issues requires more than just efficient scheduling; it demands a strategic bridge between the building's stationary energy system and the dynamic mobility needs of its users. 
While prior work addresses these challenges in isolation, a key frontier remains: the {co-optimization} of the real-time charging control policy with a strategic, user-facing negotiation policy. This requires a framework that both makes optimal control decisions and generates fair, personalized incentives. While methods like Reinforcement Learning (RL) have shown promise for handling uncertainty, the co-optimization task particularly benefits from approaches that can explicitly handle hard constraints and adapt to new scenarios without extensive re-training. For this reason, we posit that the explicit constraint handling and computational efficiency of Model Predictive Control (MPC) provide a more direct and scalable solution~\cite{shapiro2021lectures,saleem2024stochastic, solar_mpc}. An MPC-based approach offers robust uncertainty handling through sampling, which we show is comparable to an RL baseline, establishing MPC as the ideal engine for our framework.

Leveraging this MPC foundation, we introduce our negotiation framework, \textsc{CONSENT}, which directly addresses the shortcomings of prior work~\cite{liu2025, ghosh2017menu, bae2021inducing}. To achieve incentive alignment, users are presented with tailored incentives for voluntary adjustments to their requested state-of-charge (SoC) or departure time, grounded in empirical survey data. To prevent strategic misreporting, our mechanism includes formal guarantees of strategy-proofness, budget feasibility, and voluntary participation. By integrating this mechanism and validating with real-world data, we demonstrate that operator savings can be aligned with high user participation and satisfaction.
Our contributions include:

\begin{itemize}
    \item A strategy-proof, budget-feasible negotiation framework for V2B charging that co-optimizes building costs and user incentives via a novel method for generating and evaluating personalized options under uncertainty and encourages voluntary participation (Sections~\ref{sec:prob_formulation} and~\ref{sec:approach}).
    \item An empirical user study and behavioral model quantifying the trade-offs between charging convenience and monetary incentives in realistic scenarios (Section~\ref{sec:survey}).
    \item A comprehensive evaluation on real-world operational data, including extensive sensitivity analyses to demonstrate the framework's robustness and effectiveness (Section~\ref{sec:exps}).
\end{itemize}

\section{Related Research}
\label{sec:related_research}

Research in EV charging systems has largely advanced along two parallel streams. The first focuses on the {charging control problem}, where the goal is to optimize charging schedules to minimize operator costs like peak demand~\cite{he2022optimal, christensen2025scoping}. A wide array of methods have been developed, including rule-based heuristics and mathematical programming~\cite{nakahira2017smoothed, stankovic_EDF, tanguy2016optimization}. More recent advances use Model Predictive Control (MPC) for real-time optimization, effectively adapting to dynamic loads, renewables, and prices~\cite{di2014electric, yang2023ev}. However, this entire stream of research, including stochastic MPC~\cite{shapiro2021lectures}, typically solves the problem from the operator's perspective by assuming full system control, thereby overlooking the critical challenge of securing voluntary user participation~\cite{thran2025reserve, sen2025}.

Recognizing that user compliance cannot be guaranteed, the second stream of research focuses on the socio-technical challenge of {user negotiation and cooperation}. This work has produced incentive-based frameworks like menu-based pricing~\cite{ghosh2017menu} and iterative negotiation~\cite{wang2018ev, khan2022multi} to provide users with choices. These are often informed by behavioral models that predict user responses to inconvenience and rewards~\cite{shariatzadeh2025electric}, with a key finding being the importance of {user heterogeneity}~\cite{andrenacci2023literature, zahringer2024watt}. Despite being user-centric, these frameworks are often designed for large-scale aggregators and rarely integrated with the specific, real-time operational constraints of a single building.

The limitations of these separate streams highlight a significant gap at their intersection: the {co-optimization problem}, which requires strategically coupling real-time control with empirically-grounded, user-centric negotiation. Tackling this frontier requires advanced methods capable of handling uncertainty and complex decisions. {Reinforcement Learning (RL)} has emerged as a powerful tool for learning control policies in such stochastic environments~\cite{liu2025, sen2025}. However, for the specific co-optimization challenge, {Model Predictive Control (MPC)} presents a compelling alternative. While RL excels in exploration, the co-optimization problem benefits from MPC's ability to explicitly handle hard constraints (e.g., user choices, charger limits) and efficiently evaluate hypothetical negotiation scenarios. By using Monte-Carlo sampling, MPC can robustly manage uncertainty, establishing it as a powerful engine for a framework that must bridge control and negotiation.
Accordingly, our work addresses this co-optimization gap. We develop an MPC-based negotiation framework that operates directly on personalized user requests and is calibrated with empirical survey data, enabling voluntary and incentive-compatible cooperation at the building-user interface.

\section{Problem Formulation}
\label{sec:prob_formulation}

\begin{figure*}[htp]
    \centering
    \includegraphics[width=0.9\linewidth]{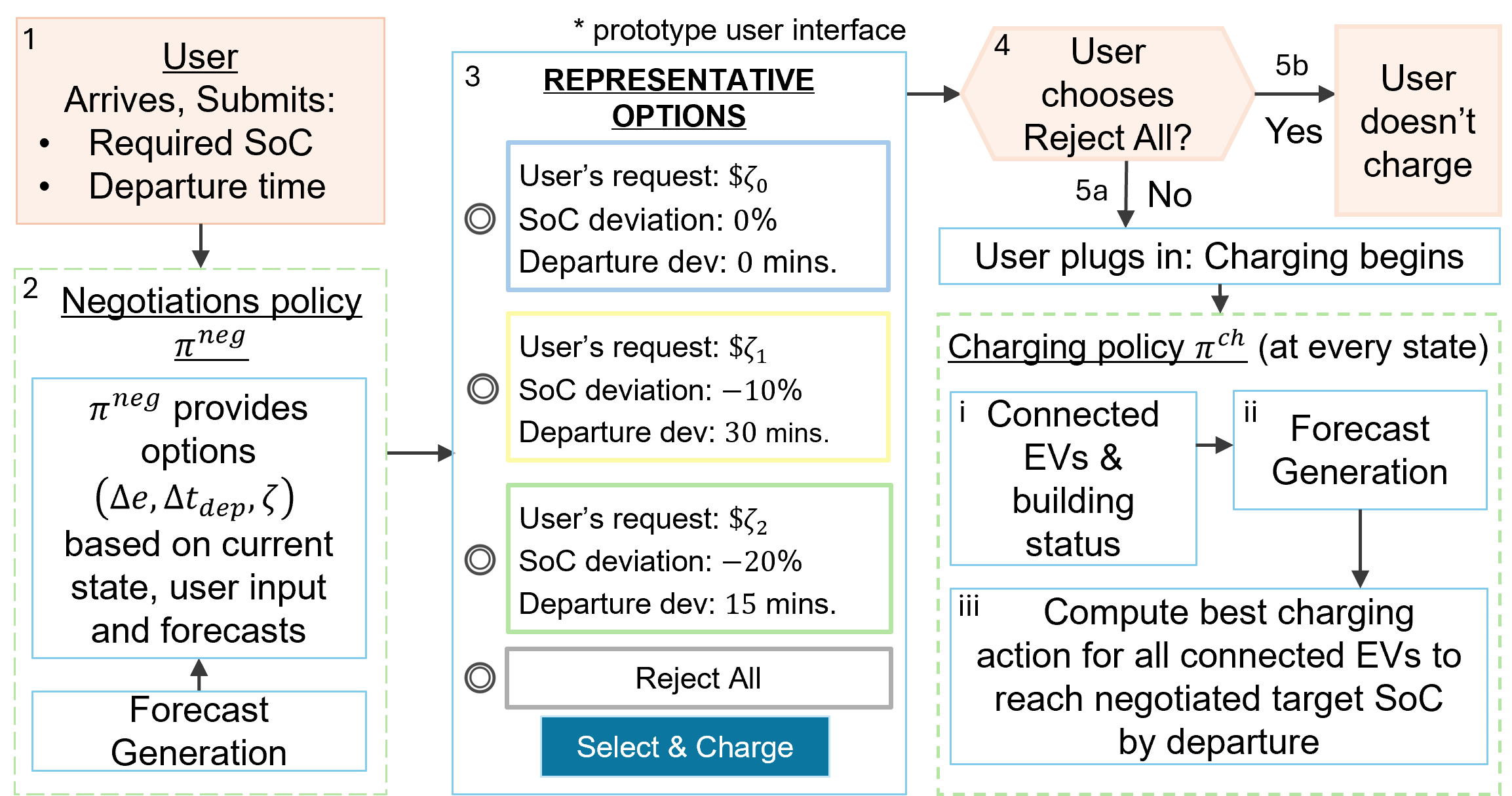}
    \caption{\textbf{Negotiation workflow: Upon arrival, users receive forecast-based offers. Accepted offers are then optimally scheduled to meet the negotiated SoC.}}
    \label{fig:user_negotiation}
    \vspace{-0.4cm}
\end{figure*}

We consider a V2B scenario where parked EVs can charge and/or discharge. The EVs follow an unknown, but regular and typically predictable pattern of arrivals, departures, and energy usage profiles. The system performs two tasks: (i) offers users cost-saving alternatives by suggesting bounded adjustments to their requested SoC and departure time; (ii) minimizes total costs by optimally scheduling EV charging/discharging under time-varying energy tariffs (\$/kWh), demand tariffs (\$/kW), charger constraints, and user-requested SoC and departure times. All EVs depart by the end of each day. A notation table is provided for reference in the Appendix~\ref{sec:appendix_a}.

{For each user, the process can be divided into three parts: \textbf{(i) Charger Allocation}: when an EV arrives, it is assigned a charger if one is available. Else, the user cannot plug-in and participate. \textbf{(ii) Charging Session Negotiation}: after plugging in, each user submits a charging session request with departure time and SoC. Based on the user's request, the system provides multiple options for target SoC and departure time, each with an associated charging cost. The user selects one of these options to begin the charging session. \textbf{(iii) Charging Actions}: During each session, the system sets charger power at every time step to minimize total cost while meeting all negotiated requirements.

\subsection{Specifications}\label{ssec:problem_spec}

\noindent{\bf Time Steps.} The optimization horizon is discretized into a finite set of time steps $\mathcal{T}$, each $\tau$ hours long.
Charging decisions are taken at the start of each time step and held constant throughout the duration.

\noindent{\bf Users and their EVs.} Let $\mathcal{V}$ be the set of EV users, each associated with a single electric vehicle (EV), and let $v$ stand for both the user and their EV. Each user $v$ has an arrival time $t_\text{arr}^v$ and a requested departure time $t_{\text{req}}^v$.
The full set of time steps when $v$ is connected is $\mathcal{T}^v \subseteq \mathcal{T}$. The EV has a battery capacity bounded by $[e_\text{min}^v, e_\text{max}^v]$, arrives with initial state of charge $e^v_\text{arr}$, and requests a target SoC at departure $e^v_{\text{req}}$, and $e_t$ provide the SoC at time $t$. The departure SoC may be revised if the negotiation leads to a different agreement.

\noindent \textbf{User Choices.}
For each user $v \in \mathcal{V}$, their choice $\theta_v \in \boldsymbol\theta$ records the arrival time $t^v_\text{arr}$, arrival SoC $e^v_\text{arr}$, negotiated departure time $\theta_v^{t_\text{dep}}$ and negotiated target SoC $\theta_v^{e_\text{dep}}$, which may be adjusted during the negotiation process, based on their requested departure time $t^v_{\mathrm{dep}}$ and $e^v_{\mathrm{req}}$ respectively. The set of all user choices within the optimization horizon is represented using $\boldsymbol\theta$.

\noindent\textbf{Negotiation Options \& Flexibility Limits.}
User flexibility is parameterized by allowable deviations in target SoC, \(\Delta e^{\mathrm{max}}_l\), and departure time, \(\Delta t^{\mathrm{max}}_{\mathrm{dep}, l}\), where \(l \in \{0, 1, \ldots, L\}\) indexes discrete flexibility levels. Each flexibility level represents a distinct negotiation option offered to users, with level \(l=0\) corresponding to no deviation from the user's originally requested departure SoC and time. The flexibility limits are determined from survey responses.

\noindent\textbf{Building's Power Draw and Energy Use.}  
Let $p^b_t \in \mathcal{P}^b$ denote the building's instantaneous power draw (kW) at time $t$. The total power draw $p_t \in \mathcal{P}$ includes both the building load and EV chargers, and the demand cost is computed from $p_t$. The building's energy cost at time $t$ is calculated as $p^b_t \cdot \tau$.

\noindent\textbf{Chargers.}  
We consider heterogeneous charger types, differing in control (controlled or uncontrolled), directionality (uni- or bidirectional), and charging rate (Level 1 or 2). Controlled chargers can be switched on/off; uncontrolled chargers draw power whenever connected to an EV. Unidirectional chargers only charge, while bidirectional chargers can also discharge. Let $\mathcal{C}$ denote all charger types with $c_k$ units of type $k$. Charging rates range within $[c_{\min}, c_{\max}]$, where $c_{\min} = 0$ for unidirectional and $c_{\min} < 0$ for bidirectional chargers. 
Charging/discharging efficiency $\eta$ models electrical losses as the ratio of output power to input power.

\noindent\textbf{Charging Rates.}  
The charging rate $ r_t^v $ (in kWh) for EV $ v $ at time $ t $ is a decision variable.

\noindent{\bf Battery Degradation.}
{The penalty term $K_{\mathrm{batt}}$ is applied to any discharging ($r^v_t<0$) to model battery degradation costs.}

\noindent \textbf{Missing SoC Cost.}
For each EV, departing below the negotiated SoC incurs a penalty of $K_{\mathbf{SOC}}$ (\$/kWh).

\noindent\textbf{Electricity Cost.}
Electricity cost comprises two components: time-varying energy charges $w_t \in \mathcal{W}$ (in \$/kWh) and demand charges $K_\text{DC}$ (in \$/kW). At each time step $ t $, total energy consumption includes the building load and EV charging. The energy cost at time $t$ is  
$
g_t = w_t \bigl(p_t^b \tau + \sum_{v \in \mathcal{V}} r_t^v \bigr),
$  
where $ p_t^b $ (kW) is the building load, $ \tau $ (hr) is the time step duration, and $ r_t^v $ (kWh) is the energy delivered to EV $ v $. The energy cost attributed to EV $ v $ sums over its connection time as  
$
g^v_t = \sum_{t \in \mathcal{T}^v} w_t \, r_t^v.
$

The building's demand cost reflects the peak power usage (kW) over a billing cycle and can contribute significantly to the total electricity costs, and are typically calculated over certain peak time periods $ t \in \mathcal{T}^{peak} \subseteq \mathcal{T} $. Peak times also incur higher energy charges, incentivizing the shift of energy consumption to off-peak times. Demand cost is computed as $K_{\mathrm{DC}} \cdot p^{\max}$, where $p^{\max} = \max_{t \in \mathcal{T}^{peak}} \left( b^p_t + \frac{1}{\tau} \sum_{v \in \mathcal{V}} r^v_t \right)$ is the maximum total power drawn at any time step in the peak periods. Demand charges require non-myopic optimization as they are non-additive and span long horizons.

\noindent\textbf{Building's Policies.}  
Let \(\pi^{neg}\) and \(\pi^{ch}\) denote the negotiation and charging policies, respectively, and define the combined policy as \(\pi = (\pi^{neg}, \pi^{ch})\). The building’s optimal policy \(\pi^* = (\pi^{neg,*}, \pi^{ch,*})\) jointly determines negotiation offers and EV charging schedules to minimize total costs, including operational expenses and user incentives, at each decision point.

\noindent \textbf{Total Cost.}  
Our objective is to minimize the building’s net expenditure over the billing period under negotiation and charging policies \(\pi\), which are controlled by the realized user choices \(\boldsymbol{\theta}\). For convenience, we omit explicit notation of this dependence, writing the cost function as
\begin{equation}\label{eq:obj}
\small
\begin{split}
& J^{\pi} (\boldsymbol{\theta} \mid \mathcal{W}) = \sum_{t \in \mathcal{T}} g_t 
+ K_{\mathbf{DC}} \cdot p^{\max} \\
& + K_{\mathbf{SOC}} \sum_{v \in \mathcal{V}} \left(e^v_{\mathrm{req}} - e^v_{\mathrm{dep}} \right)
+ K_{\mathbf{batt}} \sum_{v \in \mathcal{V}} \sum_{t \in \mathcal{T}^v} q_t^v
\end{split}
\end{equation}
where all control parameters and system actions (e.g., \(g_t, p^{\max}, q_t^{v}, e_{\mathrm{dep}}^{v}\)) are implicitly determined by the charging policy \(\pi\), itself governed by the negotiated user choices \(\boldsymbol{\theta}\). We drop these indices for clarity and readability.

\noindent\textbf{User's Marginal Utility.}
$\boldsymbol\theta_v$ denotes user $v$'s arrival and departure choices. Let $\boldsymbol{\theta}_{\setminus v}$ be the realized choices of all users who have arrived prior to $v$. For users arriving after \(v\), let \(\boldsymbol{\hat{\theta}}_{\setminus v}\) be random variables with joint distribution \(\psi\), representing all possible future arrivals along with their associated choices and flexibilities. Instead of point forecasts, the controller samples from \(\psi\) to estimate expected system costs under uncertainty, as \(\psi\) encompasses the stochastic processes of EV arrivals and user decisions (e.g., arrival/departure times and flexibility preferences).

Given the optimal policy $\pi^{*}$ (comprising both negotiation and charging decisions), user $v$'s marginal utility is defined as the expectation under $\psi$ of the cost reduction enabled by the user's participation:
\begin{equation}\label{eq:user_utility}
\small
\begin{split}
U^{\boldsymbol{\theta}_{\setminus v}, \pi^{*}}(\theta_v) = \mathbb{E}_{\boldsymbol{\hat{\theta}}_{\setminus v} \sim \psi}
    \Big[
    J^{\pi^{*}}(\boldsymbol{\theta}_{\setminus v}, \boldsymbol{\hat{\theta}}_{\setminus v} \mid \mathcal{W}')\\
    - J^{\pi^{*}}(\boldsymbol{\theta}_{\setminus v}, \theta_v, \boldsymbol{\hat{\theta}}_{\setminus v} \mid \mathcal{W}')
    \Big]
\end{split}
\end{equation}  
where the expectation integrates over all future scenarios varying in arrival numbers, user types, and their flexibility parameters. We assume the quantity \(\mathcal{W}'\) represents building loads, energy charges, and demand charge parameters.
If \(U(\theta_v) > 0\), then serving user \(v\) is expected to reduce net building costs (e.g., via peak shaving or load shifting); otherwise, if \(U(\theta_v)<0\), the user’s participation is expected to increase costs.

\noindent\textbf{User Cost.} The charging cost for user \(v\), incorporating incentives, is denoted by \(\zeta^v\). It depends on the user’s marginal utility; the following sections outline its computation.

\noindent\textbf{Solution Space.}
A feasible solution to the V2B charging and negotiation problem consists of:
\begin{itemize}
    \item \textbf{Assignment}: Each arriving EV $v \in \mathcal{V}$ is assigned to a single charger $c \in \mathcal{C}$ for its entire stay.

    \item \textbf{Negotiated User Requirements}: For each user, the system may propose adjustments to their requested SoC and departure time in exchange for an incentive, providing multiple choices of $(e^v_\text{req}, t^v_\text{dep}, \zeta^v)$ to choose from.

    \item \textbf{Charging/Discharging Schedule}: For each user's EV $v$ and each time step $t \in \mathcal{T}^v$, the charging policy $\pi^{\text{ch}}$ provides the charging rate $r^v_t$ from the feasible set determined by the assigned charger’s capabilities and efficiency. At every time step, the aggregate building and charging load must not violate power, charger, or infrastructure constraints.
\end{itemize}

The solution space includes all charger assignments, charge/discharge schedules, negotiated user requirements, user utilities that jointly satisfy system constraints, user participation guarantees, and cost minimization objectives.

\section{Approach}
\label{sec:approach}
Due to the stochastic nature of EV arrivals and the complexity of the V2B charging optimization problem, we decompose the task into two subproblems: (i) EV-to-charger assignment for newly arriving EVs and (ii) formulating the user negotiation options.

\subsection{EV-Charger Assignment \& Charging Policy}
Whenever an EV arrives, the system assigns it to an available charger using a first-come, first-served (FCFS) policy. This policy prioritizes higher-power bidirectional chargers and breaks ties by earlier departure times, ensuring fairness among users. As shown in Table~\ref{table:charger_assignment_policies} (Appendix~\ref{sec:appendix_a}), FCFS offers a simple yet effective heuristic compared to alternative strategies. After assignment, the system formulates user negotiation options.

The negotiation module presents multiple charging options to each user, each with distinct flexibility levels and corresponding user costs to encourage adaptive behavior (detailed later). Based on the user’s selected required SoC and departure time, the system applies an EV charging policy $\pi^{\text{ch}}$ to optimize charging rates for long-term building cost minimization.

The charging policy $\pi^{\text{ch}}$ employs a Monte Carlo Model Predictive Control (MC-MPC) approach that solves a mixed-integer linear program (MILP) over a short-horizon window (e.g., one day). To handle uncertainty, the MILP optimization incorporates multiple Monte Carlo–sampled scenarios of future EV arrivals and building loads. It captures key system dynamics, including state-of-charge (SoC) evolution, while minimizing a total building cost function comprising energy costs, long-term demand charges, and penalties for unmet charging requirements. This rolling-horizon formulation enables robust, near-optimal decisions that dynamically adapt to real-world stochastic conditions. The detailed formulation of the charging policy is provided in Appendix~\ref{sec:charging_policy}.

\subsection{Formulating the User Negotiation Options} 
We model the EV user negotiation process as a sequential decision-making framework triggered by each new user arrival, as illustrated in Fig.~\ref{fig:user_negotiation}.  
The objective is to design a negotiation policy that interacts with users based on their requested charging requirements and departure times, offering alternative options with associated costs to encourage more flexible charging behaviors.  
This process operates alongside an underlying charging policy, \(\pi^{ch}\), which determines each charger's rate at fixed intervals (e.g., every 15 minutes).  
The negotiation policy is computed through the following three steps.  
\noindent\textbf{Step 1: Option Generation.}  
Upon each user arrival, the system generates \(L\) charging options, each defined by distinct flexibility bounds and corresponding costs.  

\noindent\textbf{Step 2: Value Estimation.}  
For each option, a forward simulation estimates the expected user cost, considering building load forecasts, previously arrived EVs, and predicted future arrivals.  
This evaluation accounts for all possible deviations within the option’s specified limits over the decision horizon.  

\noindent\textbf{Step 3: Option Provision.}  
The system then presents the user with the option that maximizes the expected reward within the allowed deviation range, balancing building cost savings, incentive payments, and user inconvenience.

\label{sec:MDP}
We formulate user negotiation as a Semi-Markov Decision Process (SMDP):
$(\mathcal{S},\mathcal{A},\mathcal{P},r)$, 
where decision epochs occur at the arrival of each new EV. Let $t_k$ denote the arrival time of the $k$-th EV, and $\Delta t_k = t_{k+1} - t_k$ the random wait until the next EV arrives.

\noindent\textbf{State.}  
At the arrival of EV \(v\) during the \(k\)-th decision epoch, the system state $s_k$ is defined as
\[
\begin{split}
s_{k} = \bigl( t_k, \; p^b_{t_k}, \; \{ e^v_{t_k} \mid v \in \mathcal{V}_k \}, \; (e^v_\mathrm{arr}, e^v_\mathrm{req}, t^v_\mathrm{req}), p^{\mathrm{past}}_{\mathrm{max}} \bigr)
\end{split}
\]
where \(t_k\) is the current time step, \(p^b_{t_k}\) is the building’s load at time \(t_k\), \(\{ e^v_{t_k} \mid v \in \mathcal{V}_k \}\) represents the states of charge of all EVs currently connected, and \((e^v_\mathrm{arr}, e^v_\mathrm{req}, t^v_\mathrm{req})\) are the newly arrived EV’s initial SoC, requested SoC, and requested departure time, respectively. The highest power draw till epoch $k$ is recorded in $p^{\mathrm{past}}_{\mathrm{max}}$.

\noindent\textbf{Action.}  
At the \(k\)-th decision epoch, when a new user \(v\) arrives in state \(s_k\), the overall action set \(\mathcal{A}(s_k)\) comprises multiple candidate charging options organized by flexibility levels \(l \in \{1, \ldots, L\}\). Each flexibility level defines a range of allowable deviations in target SoC and departure time.
For each flexibility level \(l\), there exists a subset of actions
$
\mathcal{A}_l = \left\{ a_i = \bigl(\Delta e_i,\, \Delta t_{\mathrm{dep}, i},\, \zeta_i \bigr) \right\} \subseteq \mathcal{A}(s_k),
$
where the specific deviations \(\Delta e_i \leq \Delta e^{\mathrm{max}}_l\) and \(\Delta t_{\mathrm{dep}, i} \leq \Delta t^{\mathrm{max}}_{\mathrm{dep}, l}\) vary within the \(l\)-th deviation limits, and the associated user cost \(\zeta_i\) depends on the current state $s_k$ and elecricity pricing.
The complete action space is the union of all such subsets, \(\mathcal{A}(s_k) = \bigcup_{l=1}^{L} \mathcal{A}_l\). Candidate actions are evaluated via the reward function to select the one that best balances user deviations and charging costs per deviation limit, yielding $L$ negotiation options.

Within the $L$ options, the action set always includes a ``No deviation'' option ($l=0$) which maintains the user’s exact SoC and departure time requests, and costs \(\zeta_0\). In addition to the $L$ options, a ``Reject all'' choice lets the user opt out of facility charging and instead charge externally (e.g., at home or public stations). It allows flexibility to reject offers with unfavorable costs or deviations.

\noindent\textbf{State Transition.}  
The state transition function \(\mathcal{P} : \mathcal{S} \times \mathcal{A} \rightarrow \mathcal{S}\) models system evolution as users respond to charging offers and charging unfolds. User requests update upon choice (e.g., modified SoC or departure), and connected EVs are charged per the policy \(\pi^{ch}\), which adapts rates to changing conditions. Details of the charging policy \(\pi^{ch}\) are in the Appendix~\ref{sec:charging_policy}.

As EVs depart on schedule, chargers free up, and the system state updates to reflect the connected EVs and their SoCs, charger availability, building load, and any new arrivals, enabling accurate tracking of charging and user decisions. The SMDP state transition kernel \(\mathcal{P}(s_{k+1}, \Delta t_k \mid s_k, a_k; \pi^{ch})\) combines:  
\textbf{(i)} the stochastic EV inter-arrival distribution; and  
\textbf{(ii)} the deterministic evolution under charging policy \(\pi^{ch}\) during \(\Delta t_k\).
We do not explicitly define the state-action transition function, as our simulator leverages data-driven generative models trained on actual operational records~\cite{10595624}.

\noindent\textbf{Reward Function.}  
At each decision epoch \(k\), the reward for action \(a_i \in \mathcal{A}_l \subseteq \mathcal{A}(s_k)\) is the marginal utility \(U\) associated with \(a_i\) in the negotiation option $l \in L$. This utility quantifies the impact of the chosen action on the total system cost, conditioned on the fixed choices of other connected EVs \(\boldsymbol{\theta}_{\setminus v}\) and the predicted behavior of future users \(\boldsymbol{\hat{\theta}}_{\setminus v}\). Formally,
$
r(s_k, a_i) = U^{\boldsymbol{\theta}_{\setminus v}, \pi^*}(a_i),
$
where \(\pi^*\) denotes the joint optimal negotiation and charging policy. The reward represents the utility of user \(v\), and through this user-centric metric, it facilitates principled optimization of the negotiation policy in the sequential decision-making process.

\subsection{Negotiation Policy}

We operationalize the negotiation problem by solving the negotiation SMDP using an online, finite-horizon optimization at each decision epoch $k$ (i.e., the arrival of a new EV).

\noindent \textbf{User Cost}: For each user $v$, the cost $\zeta^v$ is the energy cost (based on time-of-use price $w_t$ and consumption) minus the utility gained from selecting the negotiation option.
\begin{equation}\label{eq:user_cost}
\zeta^v = \sum_{t \in \mathcal{T}^v} g^v_t w_t - \alpha \cdot U({\theta}_v)    
\end{equation}
where $g^v_t$ is the energy charged to the EV $v$ at time $t$ (charging is positive energy use, and discharging is negative energy use), $w_t$ is the electricity price at time $t$, and $\sum_{t \in \mathcal{T}^v} g^v_t w_t$ is the user's energy cost to be paid to the power utility. The user’s utility \(U(\boldsymbol{\theta}_v)\) can be scaled by \(\alpha\) to set the share of utility passed from the building operator to the user.

Our negotiation engine selects incentive-backed charging offers for each arriving by integrating a Monte Carlo sampling based model predictive control (MC-MPC). Our approach belongs to the family of stochastic Model Predictive Control (SMPC) methods \cite{MA2016}. Algorithm~\ref{alg:negotiation_policy} presents the user negotiation process. 

\noindent
\textbf{MC-MPC Controller with Deviation Limits.}  
The MC-MPC controller implements the negotiation policy $\pi^{neg}: (s_k, \mathcal{F}) \mapsto \mathcal{A}(s_k)$, where $s_k$ is the current state and $\mathcal{F}$ denotes a set of Monte Carlo-sampled future scenarios. Each scenario simulates anticipated EV arrivals, departures, and building loads over the prediction horizon. To account for user flexibility, the controller systematically evaluates each flexibility level $l \in \{0,1, \ldots, L\}$, where each $l$ specifies allowable bounds for SoC and departure time deviations.

For each flexibility level \(l\), the controller explores the corresponding continuous action subset \(\mathcal{A}_l\) (from the SMDP) and solves a Mixed Integer Linear Program (MILP) implementing a Sample Average Approximation ~\cite{shapiro2021lectures}. The MILP minimizes the average total cost over the sampled scenarios \(\mathcal{F}\), yielding an optimal candidate action \(a_l^*\) for each \(l\):
\begin{equation}
\small
\begin{aligned}
a^*_l = \min_{\mathcal{A}_l} \quad & \frac{1}{|\mathcal{F}|} \sum_{f \in \mathcal{F}} \Bigg[
    \sum_{i=0}^{H(t)} g_{i,f} 
    + K_{\mathrm{DC}} \cdot p^{\max}_f \\
    & + K_{\mathrm{SOC}} \sum_{v \in \mathcal{V}^f} z^v_f 
    + K_{\mathrm{batt}} \sum_{v \in \mathcal{V}} \sum_{t \in \mathcal{T}^v} q_{t,f}^v 
\Bigg]
\end{aligned}
\end{equation}
where for each sample $f$, $g_{i, f}$ is the energy cost, $p_{\text{max}, f}$ is the peak power draw, $z^v_f$ is any unmet SoC by departure for EV $v$, and $q^v_{t,f}$ is the sum of energy discharged.

The solution is subject to user deviation constraints corresponding to level $l$, operational constraints, and realized or fixed past user choices. The policy also offers the additional \textit{Reject all} choice, which uses a fixed market-rate price of the nearest available charger.
This approach enables the negotiation policy to offer robust options, striking a balance between user flexibility and cost minimization, while accounting for realistic uncertainty about future system dynamics.

\subsection{Theoretical Guarantees}

\begin{theorem}[Strategy-Proofness for Departure Time and Requested SoC]
The mechanism is strategy-proof with respect to departure time and requested SoC: a user cannot increase their utility by misreporting an earlier departure time or a higher required SoC \(\bar{\theta}_v\) than their true preferences \(\theta_v\). Formally, truthful reporting always maximizes user utility, i.e., \(U(\theta_v) \geq U(\bar{\theta}_v)\).
\end{theorem}
\noindent\emph{Proof.}
Proof is in the Appendix~\ref{sec:proofs} due to brevity of space.

\begin{theorem}[Budget Feasibility]
Any payment $\zeta^v$ to user $v \in \mathcal{V}$ must satisfy $\zeta^v \ge -U(\theta_v)$ to ensure that the building’s total cost after incentives does not exceed the baseline.\end{theorem}

\noindent\emph{Proof.} Proof is in the Appendix~\ref{sec:proofs}.

\begin{theorem}[Voluntary Participation]
For all rational users $v$, the user choice mechanism guarantees ${Y}^v_{l} \ge 0$. Thus, no user is worse off by participating, and some benefit strictly if $U(\theta_v) > 0$.
\end{theorem}

\noindent\emph{Proof.} Proof is in the Appendix~\ref{sec:proofs}.

\SetCommentSty{textit} %
\newcommand\mycommfont[1]{\textcolor{blue}{\textit{#1}}}
\SetCommentSty{mycommfont}
\begin{algorithm}[ht]
\small
\caption{Online V2B Negotiation Policy}
\label{alg:negotiation_policy}
\KwIn{Current state $S_k$; number of future samples $N$; number of flexibility levels $L$}
\KwOut{Sequence of negotiated choices $\Theta = \{\theta_1, \theta_2, \dots \}$}

\For{each new EV arrival $k = 1, 2, \dots$}{
    Observe the current system state $S_k$ at arrival time $t_k$\;
    
    Future trajectory set $\mathcal{F} \leftarrow \emptyset$\;
    \For{$i = 1$ \KwTo $N$}{
        Sample a future trajectory $f$ from the generative model given $S_k$\;
        Add $f$ to $\mathcal{F}$\;
    }
    
    Option set $\mathcal{L}_k \leftarrow \emptyset$\;
    \For{each flexibility level $l = 0$ \KwTo $L$}{
        Solve the MILP (Eq. (4)) over scenarios $\mathcal{F}$ to find the optimal option $a^*_l$ for level $l$\;
        Add $a^*_l$ to the option set $\mathcal{L}_k$\;
    }
    
    Present the option set $\mathcal{L}_k$ (plus the 'Reject all' choice) to the arriving user $v_k$\;
    Receive the user's choice $\theta_k \in \mathcal{L}_k \cup \{\text{reject}\}$\;
    
    Update EV $v_k$'s requirements based on the negotiated choice $\theta_k$\;
    
    Apply the online charging policy (Algorithm~\ref{alg:MP-MPC} in Appendix) for all time steps between $t_k$ and the next arrival $t_{k+1}$ to evolve the system to state $S_{k+1}$\;
}
\end{algorithm}

\section{Survey and User Modeling}\label{sec:survey}

\noindent

To quantify user flexibility, we conducted an anonymized stated-preference survey with 28 university participants. Respondents specified both their maximum tolerable deviations in departure time and SoC, as well as the minimum compensation required (their perceived inconvenience cost) to accept specific deviation scenarios in departure $(\Delta t_{\mathrm{dep}})$, and requested SoC $(\Delta e)$.

\noindent
(i)~\textit{Deviation limits for negotiation options:}
Survey responses (in the Appendix~\ref{sec:survey_full}) are used to group users into \(L\) negotiation options, indexed by $l$,  each with calibrated upper bounds on SoC and departure time deviations,  \(\Delta e^{\max}_l\) and \(\Delta t^{\max}_l\). These bounds establish the menu of negotiation options that the system offers. 

\noindent
(ii)~\textit{User choice and inconvenience cost model:}  
Each survey respondent \(v\) provides data across $j$ scenarios, each characterized by deviations \(\Delta e^{v}_j\) and \(\Delta t^{v}_{\mathrm{dep}, j}\), along with a self-reported required discount, which we are terming as ``inconvenience'' \(I^{v}_j\). These observations are used to estimate inconvenience sensitivities per respondent through the linear model, 
$
I^{v}_j = w^{v}_1\, \Delta e^{v}_j + w^{v}_2\, \Delta t^{v}_{\mathrm{dep}, j}.
$
Thus, for no deviation, $I^v_j=0$.
To generalize across the population and reduce dimensionality, respondents \(v\) are clustered into \(N\) user types, indexed by \(n\), each characterized by average inconvenience weights \((w^{n}_1, w^{n}_2)\).
For each user type \(n\), the estimated inconvenience cost associated with negotiation option \(l\) is computed as
$I^{n}_l = w^{n}_1\, \Delta e_{l} + w^{n}_2\, \Delta t_{\mathrm{dep}, l}$
where \(\Delta e_{l}\) and \(\Delta t_{\mathrm{dep}, l}\) represent the deviations in SoC and departure time offered for option \(l\).

User acceptance during negotiation is modeled via a logit-based choice function~\cite{bae2021inducing}. The utility of acceptance for option \(l\) by user type \(n\) for charging cost $\zeta_l$ and given an external charging cost $\bar{E}$, is
\begin{equation}\label{eq:user_satisfaction}
\small
Y^{n}_l = \Big[ \bar{E} - (\zeta_l + I^{n}_l) \Big]
\end{equation}

The probability that  user of type \(n\) accepts option \(l\) is
$
P^{n}_l = \frac{\exp(Y^{n}_l) + \varepsilon}{\sum_{l=0}^L \exp(Y^{n}_l)},
$
with a small noise term, \(\varepsilon\), for numerical stability.

This framework enables the system to estimate acceptance likelihoods personalized by inferred user types and their characteristic inconvenience costs, allowing design and optimization of incentive-aware negotiation menus aligned with user preferences. Summary statistics of the survey and clustering results are in the Appendix~\ref{sec:survey_analysis}.

\begin{table*}[t]
\centering
\small
\setlength{\tabcolsep}{8pt} %
\begin{tabular}{llrrrr@{}}
\toprule
& \textbf{Policy Description} & \textbf{Building Cost} & \textbf{User Cost} & \textbf{Monthly User Cost} & \textbf{Reject (\%)} \\
& & \textbf{(\$/month)} & \textbf{(\$/kWh)} & \textbf{(\$/month)} & \textbf{} \\
\hline
\multicolumn{6}{@{}l}{\textit{Reference Baseline}} \\
1 & Building-Only Reference & $8484 \pm 2185$ & \textemdash & \textemdash & \textemdash \\
\hline
\multicolumn{6}{@{}l}{\textit{Non-Negotiation Baselines}} \\
2 & Uncoordinated Charging~\cite{al2021slow} + Free Cost & $9005 \pm 2184$ & $0.00$ & \textemdash & \textemdash \\
3 & Smart Charging + Free Cost~\cite{liu2025} & $8785 \pm 2144$ & $0.00$ & \textemdash & \textemdash \\
4 & Smart charging + Fixed User Cost~\cite{svp} & $8457 \pm 2144$ & $0.178$ & $328 \pm 61$ & \textemdash \\
5 & Smart Charging + Utility-based Cost & \bm{$8413 \pm 2183$} & $0.183$ & $372 \pm 64$ & \textemdash \\
\hline
\multicolumn{6}{@{}l}{\textit{Negotiation-Based Policies}} \\
6 & Menu-based Negotiation~\cite{bae2021inducing, ghosh2017menu} & $8479 \pm 2153$ & $0.157$ & $309 \pm 63$ & 34.89 \\
7 & \textbf{CONSENT (Our Approach)} & $8480 \pm 2171$ & \textbf{0.138} & \textbf{254 $\pm$ 55} & \textbf{21.99} \\
\hline
\multicolumn{6}{@{}l}{\textit{Ablation Study}} \\
8 & \textsc{CONSENT} (without Reject Option) & $8461 \pm 2155$ & $0.165$ & $299 \pm 60$ & \textemdash \\

\bottomrule
\end{tabular}
\caption{Comparative Evaluation of V2B Policies. The table progresses from a building-only reference to increasingly sophisticated baselines, culminating in our proposed negotiation framework (\textsc{CONSENT}). Lower values are better for all the metrics.}
\label{tab:combined_results}
\vspace{-0.3cm}
\end{table*}

\section{Experiments}\label{sec:exps}

We evaluate our negotiation framework using a modular simulator~\cite{10595624} of a building with multiple chargers, dynamic energy and demand charges, and realistic EV arrival and departure patterns. It captures stochastic user preferences, building load forecasts, and charger constraints, supporting online or ahead-of-time scheduling.

\noindent\textbf{Hardware.} Experiments ran on a 32-core, 6.2 GHz, 128 GB RAM machine using IBM CPLEX~\cite{ibm_cplex} for MILP.

\noindent\textbf{Data Collection.}  \label{sec:data_gen}
We use data from a commercial site in Santa Clara, California (May 2023 – November 2024), including building power usage, charger activity, and vehicle telemetry. All EVs used support bidirectional charging, and charging is modeled with a three-segment piecewise linear SoC curve.
We sampled $100$ such episodes, evenly split into $50$ for future sampling and $50$ for testing. 
Only weekdays are analyzed, as employee presence and demand charges are negligible on weekends. Figures~\ref{fig:car_arrival_departure} and~\ref{fig:soc} (Appendix~\ref{sec:gen_models}) show EV arrival/departure and SoC distributions.
The site has controllable, 10 unidirectional (0–20 kW) and 5 bidirectional (±20 kW) chargers. Silicon Valley Power's time-of-use and demand charge rates~\cite{svp} define \(\mathcal{W}\) where peak hours (6 AM to 10 PM) are $\$0.178$/kWh and off-peak hours are $\$0.137$/kWh. Demand charge \(K_{\mathrm{DC}}\) is $\$11.67$/kW.
The penalty for missing SoC is set at $K_{\mathrm{SOC}} = \$0.5$/kWh, and battery degradation penalty is set to $K_{\mathrm{batt}} = \$0.05$/kWh for discharging. The external charging cost is set to $\bar{E} = \$0.30$/kWh, reflecting the average residential EV charging rate in the area, including amortized charger installation costs. More details in the Appendix~\ref{sec:gen_models}. 

\noindent\textbf{Evaluation Metrics.}  
Each scenario is evaluated using the following metrics.  
\textit{\textbf{Building cost}} measures the total monthly cost incurred by the operator.  
\textit{\textbf{User cost}} is the average cost per kilowatt-hour delivered to users, including incentives. 
\textit{\textbf{Monthly user cost}} denotes cumulative user expenditure over the month.  
\textit{\textbf{Reject (\%)}} indicates the fraction of users unable or unwilling to charge at the building.

\noindent\textbf{Forecasting.} 
EV arrivals, departures, and SoC requests are forecasted with generative models trained on nearly two years of telemetry. 
The building operator confirmed EV schedules are independent of building load and past charging, allowing decoupled pre-sampling. Detailed EV behavior and building load models, with performance metrics, are reported in the Appendix~\ref{sec:gen_models}.
For experimenting, we sample $\mathcal{F} = 10$ future samples for $\pi^{\text{neg}}$ and $\mathcal{F} = 30$ for $\pi^{\text{ch}}$.

\noindent\textbf{Negotiation Options.}  
Based on the user survey, we define the negotiation options used in the experiment.  
Each user is presented with \(L = 3\) alternatives in addition to their exact request (\(l = 0\)).  
Each option is represented as a tuple \((\Delta e^{\max}_l, \Delta t^{\max}_{\mathrm{dep},l})\), indicating the maximum deviations in SoC and departure time.  
Specifically, option \(l=0\) is \((0, 0)\), option \(l=1\) is \((6.25\%, 30~\text{min})\), option \(l=2\) is \((10\%, 15~\text{min})\), and option \(l=3\) is \((20\%, 105~\text{min})\).  
An additional ``Reject all'' option allows users to opt out of charging at the building.

\noindent\textbf{User Types.}  
Survey analysis (Appendix~\ref{sec:survey_analysis}) identifies $N=4$ user types, defined by time and SoC sensitivity.  
Without ground-truth labels, users are assigned types based on survey proportions.
\begin{center}
\centering
\small
\label{tab:user_type_stats}
\begin{tabular}{c c c c}
\
{Type} & {SoC Sensitivity ($w_1$)} & {Time Sensitivity ($w_2$)} & {Share} \\
\midrule
i & $0.0489 \pm 0.02$ & $0.1250 \pm 0.01$ & 10.7\% \\
ii & $0.0133 \pm 0.01$ & $0.0346 \pm 0.02$ & 53.6\% \\
iii & $0.0362 \pm 0.01$ & $0.0673 \pm 0.01$ & 32.1\% \\
iv & $0.0000 \pm 0.00$       & $0.1083 \pm 0.00$         & 3.6\% \\
\hline
\end{tabular}
\end{center}

\noindent\textbf{Baseline Policies.} We evaluate our framework against a series of progressively sophisticated baselines detailed in Table~\ref{tab:combined_results}. The analysis begins with the \textit{\textbf{Building-Only}} reference, which establishes the ideal operational cost without any EV load. We then model the common real-world scenario of \textit{\textbf{Uncoordinated Charging}}~\cite{al2021slow}, where EVs charge for free upon arrival. Building upon this, the \textit{\textbf{Smart Charging + Free Cost}} policy~\cite{liu2025} isolates the benefit of centralized coordination by introducing an intelligent scheduling algorithm ($\pi^{\text{ch}}$) while keeping charging free for users. To introduce a simple economic model, we evaluate the \textit{\textbf{Smart Charging + Fixed Cost}} policy, which sets the user price based on the local utility's peak energy rate~\cite{svp}. Advancing this, the \textbf{Smart Charging + Utility-based Cost} policy represents a purely operator-centric model that pairs the scheduling algorithm with our utility based pricing function (Eq.~(\ref{eq:user_cost})) to maximize building savings. Finally, to benchmark against state-of-the-art interactive systems, we include the \textit{\textbf{Menu-based Negotiation} }policy~\cite{bae2021inducing, ghosh2017menu}, which introduces user choice and the ability to reject offers, serving as a direct precursor to our proposed framework.

\section{Results}

\noindent We evaluate our framework, \textsc{Consent}, which integrates smart charging, utility-based cost generation tailored to user flexibility, and the option for users to reject offered choices. Its performance is compared against a series of progressively complex baseline policies, with results summarized in Table~\ref{tab:combined_results}. For reference, the building’s operational cost without any EV charging load (Policy~1) is nearly \$8{,}484 per month, representing an ideal peak-shaving benchmark.

\noindent\textbf{Baseline Policy Performance.} 
The real-world baseline of uncoordinated free charging (Policy~2)~\cite{al2021slow} raises the building’s monthly cost by over 6\%.  
A smart scheduling algorithm (Policy~3)~\cite{sen2025, liu2025} mitigates this increase but still costs more than the no EV reference.  
Introducing a simple, fixed price for users (Policy~4) based on the utility's peak energy pricing~\cite{svp} further reduces the building's cost, bringing it nearly to the ideal no EV baseline. Finally, using our utility-based, demand-aware user pricing (Policy~5) enables the building to achieve its {lowest operational cost} among all policies. However, this purely operator-centric optimization leads to the {highest per-unit user cost}, leading to negotiation-based approaches.

\noindent\textbf{Negotiation Framework Performance.} 
Building on these baselines, our first negotiation-based comparison is Policy~5, which implements a state-of-the-art, menu-based approach consistent with prior work~\cite{ghosh2017menu, bae2021inducing}.  
While this method reduces user costs, it exhibits a high rejection rate of nearly 35\%, indicating misaligned incentives.  

Our proposed framework, \textsc{CONSENT} (Policy~6), strikes the best overall balance between operator and user objectives, and offers a \textbf{12\% lower user cost} at \$0.138/kWh and substantially reduces the rejection rate from 34.89\% to \textbf{21.99\%}, compared to {Menu-based Negotiation}. This user cost is over 61\% lower than average public charging rates in the United States~\cite{ev_price}. Simultaneously, the framework reduces building operator costs by 5.83\% compared to uncoordinated charging and by nearly 3.5\% compared to an optimized smart-charging policy with free user access (Policy~3).

To isolate the importance of user choice, we performed an ablation study by removing the reject option (Policy 7). The results show that while forcing participation slightly reduces the building's cost to \$8461/month, it does so at the expense of a significantly higher user cost (\$0.165/kWh). This trade-off highlights the critical role of providing user autonomy; the reject option is essential for achieving the most favorable outcome for users.
Generating all options for a user takes $\sim$0.8s on average, with each charging option computation requiring $\sim$0.99s. Our method is scalable over larger problem instances, as shown in Figure~\ref{fig:runtime_comparison}.

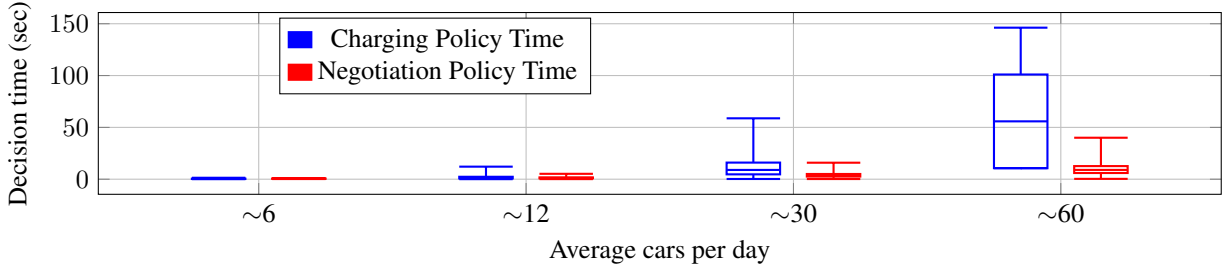
\begin{figure}[th]
\centering
\begin{tikzpicture}
\begin{axis}[
    boxplot/draw direction=y,
    ylabel={Decision time (sec)},
    xtick={1,2,3,4},
    xticklabels={$\sim$6, $\sim$12, $\sim$30, $\sim$60},
    xlabel={Average cars per day},
    width=\columnwidth,
    height=4cm,
    grid=major,
    legend style={
        at={(0.3, 0.55)},
        anchor=south,
        legend columns=1
    },
    boxplot/box extend=0.2,
    boxplot/whisker extend=0.2,
    every boxplot/.style={thick},
    every median/.style={ultra thick, black},
    legend image code/.code={
        \draw[#1, fill=#1] (0cm,-0.1cm) rectangle (0.3cm,0.1cm);
    }
]

\addplot[
    blue, solid,
    boxplot prepared={
        median=0.3636, upper quartile=0.5368, lower quartile=0.2286,
        upper whisker=1.3813, lower whisker=0.0564
    },
    boxplot/draw position=0.85
] coordinates {};
\addlegendentry{Charging Policy Time}

\addplot[
    red, solid,
    boxplot prepared={
        median=0.4977, upper quartile=0.6281, lower quartile=0.3899,
        upper whisker=1.0662, lower whisker=0.1621
    },
    boxplot/draw position=1.15
] coordinates {};
\addlegendentry{Negotiation Policy Time}

\addplot[
    blue, solid,
    boxplot prepared={
        median=1.4145, upper quartile=2.3353, lower quartile=0.7957,
        upper whisker=12.1020, lower whisker=0.0485
    },
    boxplot/draw position=1.85
] coordinates {};
\addplot[
    blue, solid,
    boxplot prepared={
        median=8.9360, upper quartile=16.0235, lower quartile=4.6831,
        upper whisker=58.7501, lower whisker=0.2067
    },
    boxplot/draw position=2.85
] coordinates {};
\addplot[
    blue, solid,
    boxplot prepared={
        median=55.824, upper quartile=101.013, lower quartile=10.635,
        upper whisker=146.202, lower whisker=10.635
    },
    boxplot/draw position=3.85
] coordinates {};

\addplot[
    red, solid,
    boxplot prepared={
        median=1.2926, upper quartile=1.7199, lower quartile=0.9172,
        upper whisker=5.1955, lower whisker=0.1014
    },
    boxplot/draw position=2.15
] coordinates {};
\addplot[
    red, solid,
    boxplot prepared={
        median=3.6480, upper quartile=5.0260, lower quartile=2.5456,
        upper whisker=15.9125, lower whisker=0.1685
    },
    boxplot/draw position=3.15
] coordinates {};
\addplot[
    red, solid,
    boxplot prepared={
        median=8.9640, upper quartile=12.6481, lower quartile=5.9741,
        upper whisker=40.0112, lower whisker=0.3259
    },
    boxplot/draw position=4.15
] coordinates {};

\end{axis}
\end{tikzpicture} 
\vspace{-0.3cm}
\caption{
Comparison of computational runtime per decision across varying EV arrival intensities.
}
\label{fig:runtime_comparison}
\end{figure} 

\begin{table}[th]
\centering
\small
\setlength{\tabcolsep}{8pt} %
\begin{tabular}{l c c c c@{}}
\toprule
\textbf{User Profile} & \textbf{Bldg. Cost} & \textbf{User Cost} & \textbf{Monthly User} & \textbf{Reject} \\
\textbf{Scenario} & \textbf{(\$/mo)} & \textbf{(\$/kWh)} & \textbf{Cost (\$/mo)} & \textbf{(\%)} \\
\midrule
\multicolumn{5}{@{}l}{\textit{\textsc{CONSENT} (with Reject option)}} \\
Baseline (Survey) & $8480 \pm 2171$ & $0.138$ & $254 \pm 55$ & $21.99$\\
More Flexible (↓25\% $w_1$,$w_2$) & $8311 \pm 2098$ & $0.159$ & $293 \pm 63$ & $19.33$ \\
Range-Sens. (↑50\% $w_1$) & $8311 \pm 2120$ & $0.156$ & $287 \pm 62$ & $23.09$ \\
Time-Sens. (↑50\% $w_2$) & $8395 \pm 2151$ & $0.155$ & $285 \pm 61$ & $22.87$ \\
Less Flexible (↑25\% $w_1$,$w_2$) & $8492 \pm 2171$ & $0.166$ & $306 \pm 66$ & $22.65$ \\
\midrule
\multicolumn{5}{@{}l}{\textit{\textsc{CONSENT} without Reject option}} \\
Baseline (Survey) & $8461 \pm 2155$ & $0.165$ & $304 \pm 50$ & \textemdash\\
More Flexible (↓25\% $w_1$,$w_2$) & $8349 \pm 2100$ & $0.195$ & $359 \pm 59$ & \textemdash\\
Range-Sens. (↑50\% $w_1$) & $8485 \pm 2110$ & $0.188$ & $346 \pm 57$ & \textemdash\\
Time-Sens. (↑50\% $w_2$) & $8412 \pm 2111$ & $0.190$ & $350 \pm 57$ & \textemdash\\
Less Flexible (↑25\% $w_1$,$w_2$) & $8503 \pm 2180$ & $0.201$ & $370 \pm 61$ & \textemdash\\
\midrule
\multicolumn{5}{@{}l}{\textit{Menu-based Negotiation~\cite{bae2021inducing, ghosh2017menu}}} \\
Baseline (Survey) & $8479 \pm 2153$ & $0.157$ & $289 \pm 52$ & $34.89$\\
More Flexible (↓25\% $w_1$,$w_2$) & $8352 \pm 2121$ & $0.176$ & $324 \pm 58$ & $33.50$ \\
Range-Sens. (↑50\% $w_1$) & $8501 \pm 2119$ & $0.181$ & $333 \pm 60$ & $36.63$ \\
Time-Sens. (↑50\% $w_2$) & $8497 \pm 2101$ & $0.182$ & $335 \pm 60$ & $36.81$ \\
Less Flexible (↑25\% $w_1$,$w_2$) & $8521 \pm 2181$ & $0.193$ & $355 \pm 64$ & $36.46$ \\
\bottomrule
\end{tabular}
\caption{Sensitivity of negotiation outcomes to different user flexibility profiles where we compare our framework, \textsc{CONSENT}, against a menu-based and an ablated baseline.}
\label{tab:sensitivity_analysis}
\vspace{-0.5cm}
\end{table}

\noindent\textbf{Sensitivity to User Flexibility Profiles.}
To evaluate the robustness of our framework, we conduct a sensitivity analysis, with results shown in Table~\ref{tab:sensitivity_analysis}. We test three policies against five simulated user populations, each representing a different flexibility profile: a baseline derived from our survey, populations that are generally more or less flexible, and populations specifically sensitive to delays in departure time or reductions in SoC. 
For our approach, \textsc{CONSENT}, a clear trade-off emerges. When interacting with a \textit{More Flexible} user population (lower sensitivity), the building operator can achieve greater savings, reducing costs from \$8480 to \$8311 per month. This increased system flexibility also leads to more agreeable offers, lowering the rejection rate from 21.99\% to 19.33\%. However, the average user cost per kWh increases, as the system optimizes for its own objectives. Conversely, a \textit{Less Flexible} population limits optimization opportunities, resulting in higher user costs and a slightly higher rejection rate. Notably, making users more sensitive to range (\textit{Range-Sensitive}) provides more cost-saving potential for the building than making them sensitive to time (\textit{Time-Sensitive}).

This analysis highlights the value of our framework. The ablation study, \textit{Our Approach without Reject option}, shows that while the building can achieve even lower costs by forcing participation, it does so at the expense of consistently higher user costs across all profiles. Furthermore, compared to the \textit{Menu-based Negotiation} baseline, our method consistently achieves a much lower rejection rate across all scenarios (e.g., 19.33\% vs. 33.50\% for more flexible users), demonstrating a superior alignment of incentives. These findings confirm that our approach is robust and effectively balances operator and user objectives across diverse user populations.
\begin{table}[htbp] 
\centering
\setlength{\tabcolsep}{8pt} %
\small
\begin{tabular}{@{}clcccccc@{}}
\toprule
\textbf{Utility Sharing} & \textbf{Building's Cost (\$/mo} & \textbf{User's Cost (\$/kWh} & \textbf{Reject(\%)} \\
\midrule
100\% sharing & $8480.11 \pm 2171.28$ & \textbf{0.138} \bm{$\pm$} \textbf{0.03} & \textbf{21.99} \\
90\% sharing & $8469.79 \pm 2173.71$ & $0.148 \pm 0.02$ & $22.14$ \\
75\% sharing & $8451.28 \pm 2171.93$ & $0.158 \pm 0.02$ & $23.78$ \\
50\% sharing & $8427.71 \pm 2163.53$ & $0.170 \pm 0.01$ & $24.06$ \\
25\% sharing & $8411.49 \pm 2155.76$ & $0.176 \pm 0.01$ & $25.33$ \\
10\% sharing & \textbf{8397.71} \bm{$\pm$} \textbf{2147.44} & $0.178 \pm 0.00$ & $26.88$ \\
\bottomrule
\end{tabular}
\caption{Performance across utility-sharing ratios $\alpha$ from user cost equation~\eqref{eq:user_cost} (100\% to 10\%): lower utility sharing lowers building cost but raises user costs and simulated rejections, highlighting trade-offs in operational savings and user acceptance.}
\label{tab:cost_sharing}
\vspace{-0.5cm}
\end{table}

\noindent \textbf{Sensitivity to Cost-Sharing Levels.}  
Table~\ref{tab:cost_sharing} summarizes the negotiation framework’s performance across cost-sharing ratios from 100\% to 10\%, by tuning \(\alpha\) in Eq.~\eqref{eq:user_cost}. As \(\alpha\) decreases, building costs fall, reaching peak savings at 10\% sharing, while user costs, satisfaction, and participation also decline. Notably, even with 90\% utility sharing, building costs drop below the building's baseline load without EVs (Policy~1 in Table~\ref{tab:combined_results}), while user costs remain lower than all other policies. This shows how strategic cost allocation drives both operational savings and user affordability, underscoring the mechanism’s robustness and adaptability.

\color{black}

\section{Conclusion}

We introduced \textsc{CONSENT}, a novel negotiation framework built on a stochastic Model Predictive Control (SMPC) foundation that acts as a strategic bridge between building energy systems and EV mobility needs. Empirically validated with real-world data, it co-optimizes operator and user objectives via a novel method to generate personalized options. This results in a 22\% user cost reduction and a significantly lower rejection rate than negotiation baselines, all while achieving higher building savings than free to the user, purely operator-centric policies, successfully transforming V2B charging into a platform for collaboration.

Our strategy-proof guarantee ensures fairness in any single negotiation by making it unprofitable to misreport needs. Future work could extend this concept from single-shot interactions to a dynamic, repeated-game setting. While our MPC-based framework adapts to new statistical patterns, a game-theoretic approach would model users as rational agents who strategically adapt their behavior over time. Developing mechanisms to anticipate these shifts and find a stable, efficient equilibrium is a key direction for future research.

\section{Acknowledgements}

This material is based upon work sponsored by Nissan Advanced Technology Center-Silicon Valley. Results presented in this paper were obtained using the Chameleon Testbed supported by the National Science Foundation. Any opinions, findings, conclusions, or recommendations expressed in this material are those of the authors and do not necessarily reflect the views of Nissan.

\bibliographystyle{unsrt} 
\bibliography{main}

\clearpage
\appendix
\section{Appendix}\label{sec:appendix_a}

\pgfplotstableread[col sep=comma]{charger_assignments_scaled.csv}{\chargerassignments}

\begin{table}[!ht]
\centering
\small
\pgfplotstabletypeset[
    create on use/bill/.style={
        create col/assign/.code={%
            \edef\entry{\thisrow{c}$\pm$\thisrow{d}}
            \pgfkeyslet{/pgfplots/table/create col/next content}{\entry}
        }
    },
    columns ={a, b, bill},
    column type=c,
    columns/a/.style={string type},
    columns/b/.style={string type,column type=c},
    columns/bill/.style={string type,column type=c},
    fixed,
    fixed zerofill,
    precision=2,
    every head row/.style={
        output empty row,
        before row={%
            \toprule
            {Assignment} & {Tie Breaker} & {Building Cost with Smart Charging (\$)}\\
        }, 
        after row=\midrule},
    every last row/.style={after row=\bottomrule},
    every row 0 column 0/.style={postproc cell content/.style={@cell content={\textbf{##1}}}}, %
    every row 0 column 1/.style={postproc cell content/.style={@cell content={\textbf{##1}}}}, %
    ] {\chargerassignments}
\caption{Charger Assignment and tiebreaker comparisons with an oracle, optimal policy (MILP). Assigning to Bidirectional chargers first and then breaking ties by assigning them to the EV that departs later results in the lowest bill. Lower is better.}
\label{table:charger_assignment_policies}
\end{table}

\begin{table*}[ht]
\centering
\footnotesize
\begin{tabular}{|l|p{11.5cm}|}
\toprule
\textbf{Variable} & \textbf{Description} \\ \toprule

$t \in \mathcal T$, $\mathcal{T}^{\text{peak}} \subseteq \mathcal{T}$ & Set of discretized time steps of uniform length $\tau$ (e.g., 0.25 hr), Set of peak periods (for demand cost) \\ \midrule

$v \in \mathcal V$ & Set of EV users in the system \\ \midrule
$[t^v_{arr}, t^v_{req}] \in \mathcal{T}^v \subseteq \mathcal{T}$ & Arrival  and required departure time of EV user $v$ \\ \midrule
$[e^v_{min}, e^v_{max}]$ & Battery capacity constraints for EV $v$ SoC \\ \midrule
$e^v_t, e^v_{arr}, e^v_{req}$ & State of charge (SoC) of user $v$'s EV at time step $t$, at arrival, and the required SoC at departure. \\ \midrule

$\theta_v = \{t^v_{arr}, e^v_{arr}, \theta_{t^{dep}}^v, \theta_{e^{dep}}^v\}$ & User parameters: arrival time, arrival SoC, departure time, departure SoC \\ \midrule
$c \in \mathcal C, c_k$ & Set of charger types, and their counts \\ \midrule
$c_{min}, c_{max}, \eta$ & Minimum and maximum charging rates supported by the charger, and its efficiency (charging/discharging) \\ \midrule
$\phi_t : \mathcal{V}_t \leftrightarrow \mathcal{C}_t$ & One-to-one mapping between user's EVs and chargers at time $t$ \\ \midrule

$p^b_t$ & Building's power load (kW) (without chargers) at time step $t$ \\ \midrule

$w_t$ & Energy price (\$/kWh) at time $t$ \\ \midrule
$K_{\mathrm{DC}}$ & Demand charge rate (\$/kW of peak demand) \\ \midrule
$K_{\mathrm{SOC}}$ & Penalty coefficient for unfulfilled SoC in objective \\ \midrule
$K_{\mathrm{batt}}$ & Penalty coefficient for battery degradation due to discharging \\
\bottomrule

\end{tabular}
\caption{Key Input Variables}
\label{tab:input_variables}
\end{table*}

\section{Charging Policy}\label{sec:charging_policy}

\begin{table*}[ht]
\centering
\footnotesize
\begin{tabular}{l|p{12cm}}
\toprule
\multicolumn{2}{l}{Charging decision variables:} \\ \midrule
\textbf{Variable} & \textbf{Description} \\ \midrule
$a^{v,c}_{t} \in \{0,1\}$ & Binary variable; 1 if EV $v$ is assigned to charger $c$ at time $t$, otherwise 0 \\ \midrule
$r^v_t$ & Charging/discharging rate (kW) for EV $v$ at time step $t$ (decision variable) \\ \midrule

$g^v_t, g_t$ & Energy draw by EV of user $v$ and the total (building + EV charging) at time $t$ \\ \midrule
$y^{v, \delta}_t$ & Indicates the current SoC segment in piece-wise linear SoC curve for EV $v$  \\ \midrule
$x^{v,c,\delta}_t$ & Links maximum charging rate to SOC segment  \\ \midrule
$p_t$ & Total power draw (building + EV charging) at time $t$ \\ \midrule
$p^{\text{max}}$ & Peak power over peak periods (for demand charges) \\ \midrule
$q^v_t$ & Auxiliary variable tracking discharge to limit battery wear \\ \midrule
$z^v$ & SoC shortfall (absolute difference between required and achieved SoC at departure) for EV $v$ \\ \midrule

\addlinespace[6pt]
\multicolumn{2}{l}{Additional Negotiation  decision variables:} \\ \midrule

$\lambda^v_t \in \{0, 1\}$ & Indicates departure time step of EV $v$ \\ \midrule
$\sigma^v_{\text{soc}}, \sigma^v_{\text{dep}}$ & Deviation from requested SoC and departure time for EV $v$ \\ \midrule
$U(\theta_v)$ & Marginal utility of user $v$ defined as cost reduction due to $v$'s participation \\ \midrule
$\zeta$ & User charging cost computed as energy cost minus shared utility \\ \midrule
$L$ & Set of generated options for negotiations \\ 
\bottomrule

\end{tabular}
\caption{Key Decision Variables}
\label{tab:decision_variables}
\end{table*}

We formulate the V2B charging policy $\pi^{ch}$ using Model Predictive Control (MPC), which utilizes a Monte-Carlo sampling and a Mixed-Integer Linear Program (MILP). The Monte-Carlo sampling provides a range of possible future trajectories in terms of EV arrivals and user requests, while the MILP aims to minimize the total cost while satisfying user charging requests and operational constraints. The model captures discrete charger assignments, models battery SoC dynamics using piecewise linear segments, and incorporates realistic electricity tariffs that include both time-of-use energy prices and demand charges. This formulation ensures that charging schedules respect user-specified arrival times, departure times, and requested SoC targets, while optimizing energy usage to flatten demand peaks and reduce total cost for the building operator.

The MPC framework enables real-time implementation in a dynamic environment with uncertain future arrivals and building loads. The MPC controller operates in a rolling horizon fashion: at each decision epoch (e.g., every 15 minutes), it solves the MILP  using up-to-date forecasts of building load, EV arrivals, and user requests. By repeatedly re-optimizing at each time step as new information becomes available, the MPC policy adapts charging and discharging actions to minimize expected cumulative cost over the remaining horizon.

\noindent\textbf{EV to Charger Assignment function.}
At each time step $t$, chargers and EVs are matched one-to-one: each charger serves at most one EV, and each EV connects to at most one charger based on the first-come-first-served (FCFS) assignment policy.The assignment policies are summarized in Table~\ref{table:charger_assignment_policies}. We select the best-performing approach: Bidirectional-first with departure time as the tie-breaker. Formally, this defines a mapping $\phi_t : \mathcal{V}_t \leftrightarrow \mathcal{C}_t$, where $\mathcal{V}_t$ and $\mathcal{C}_t$ are the sets of EVs and chargers active at $t$. If $\phi_t(v) = c$, then $\phi_t^{-1}(c) = v$, respecting infrastructure constraints.

\noindent\textbf{Charging Rates.}
In addition to the charging rate limits of the chargers, to capture the typical slowdown in charging as an EV battery fills, we model charging using piecewise-linear segments $\delta \in \Delta$. Each segment is defined by a maximum charging rate $ r_{\max}^{v,\delta} $ valid over SoC intervals $[e_{\min}^{v,\delta}, e_{\max}^{v,\delta}]$. During discharging, these piece-wise functions are not considered, and the rate of discharge is considered to be linear at all battery levels.

\noindent\textbf{Decision Variables}

\begin{itemize}
    \item $a^{v,c}_{t} \in \{0,1\}$: Binary assignment variable indicating whether electric vehicle $v$ is assigned to charger of type $c$ at time step $t$.
    \item $y^{v, \delta}_{t} \in \{0,1\}$: Indicator variable equal to 1 if the state of charge (SoC) of EV $v$ lies within segment $\delta$ of the piecewise-linear SoC curve at time $t$.
    \item $x^{v,c,\delta}_{t} \in \{0,1\}$: Binary variable coupling the SoC segment $\delta$ and charger assignment $a^{v,c}_{t}$ to regulate feasible charging rates.
    \item $r^{v}_{t} \in [c_{\min}, c_{\max}]$: Continuous variable indicating charging (or discharging) power for vehicle $v$ at time step $t$, where bounds depend on charger directionality.
    \item $e^{v}_{t} \in [e^{v}_{\min}, e^{v}_{\max}]$: Continuous variable denoting the state-of-charge (SoC) of EV $v$ at the start of time step $t$.
    \item $g_{t} \in [0, \infty)$: Energy cost incurred by building and charging load during time step $t$, constrained so that net energy recovered from vehicles never exceeds current building energy use.
    \item $q^{v}_{t} \geq 0$: Auxiliary continuous variable penalizing excessive discharging from vehicle $v$ at time $t$.
    \item $z_{v} \geq 0$: Continuous variable tracking the SoC deficit at the time of vehicle $v$'s departure; used to penalize unmet charging targets and maintain user satisfaction.
    \item $p^{\max} \geq 0$: Peak building power draw, defined as the maximum instantaneous power consumption across all time steps $\max_{t}(p_{t})$.
\end{itemize}

\noindent\textbf{Constraints}
\label{sec:constraints}
The following constraints are modeled the MILP formulation:

\noindent\emph{EV to Charger Assignment.} 
For the duration of the EV $v$, \( \mathcal{T}^v \subseteq \mathcal{T} \) we already assumed EV to charger assignment is done exogenously, and is captured by $a_{v,c,t}$.

To ensure that an EV remains continuously assigned to the same charger throughout its stay, we impose:
\begin{equation}\label{eq:car_charger}
\textstyle
\forall v \in \mathcal{V},\ \forall c \in \mathcal{C},\ \forall t \in \mathcal{T}^v \setminus \{t_{\text{arr}}^v\}: \quad
a^{v,c}_{t-1} - a^{v,c}_{t} = 0
\end{equation}

\noindent\emph{Modeling Charging Dynamics via Piecewise Linear Curves.} 
We model EV battery dynamics using piecewise-linear approximations of the charging curve, divided into \( \Delta \) SoC segments. For each EV \( v \), the SoC at any time must lie within exactly one segment which ranges between $\left[e^{v, \delta}_{min}, e^{v, \delta}_{max} \right]$ for $\delta \in \Delta$ segments, as indicated by $y^{v, \delta}_t$:

\begin{equation}\label{eq:soc_seg}
\footnotesize
\forall v \in \mathcal{V},\ \forall t \in \mathcal{T}^v,\ \forall \delta \in \Delta: \quad
\begin{cases}
e^v_t \geq e^{v, \delta}_{\min} \cdot y^{v, \delta}_t - (1 - y^{v, \delta}_t) \\
e^v_t \leq e^{v, \delta}_{\max} \cdot y^{v, \delta}_t + M_1 (1 - y^{v, \delta}_t)
\end{cases}
\end{equation}

where \( M_1 \) is a large constant, and we set it to \( e^{v}_{\max} \) to relax the bounds when a segment is inactive. Choosing a smaller value, such as choosing $M_1 = e^{v,\delta_{max}}$ would cancel the indicator variable $y^{v, \delta}_{t}$ and invalidate the constraint; we therefore retain $M_1 = e^v_{max}$.

Each EV must belong to exactly one linear segment at any time:
\begin{equation}
\textstyle
\forall v \in \mathcal{V},\ \forall t \in \mathcal{T}: \quad
\sum_{\delta \in \Delta} y^{v, \delta}_{t} = 1
\end{equation}

We define auxiliary variables \( x^{v,c, \delta}_{t} \) to link charger assignment \( a^{v,c}_{t} \) and SoC linear segment \( y^{v, \delta}_{t} \). These variables are used to limit the charging energy $r^v_t$ in each segment of the SoC curve:
\[
\forall v \in \mathcal{V},\ \forall c \in \mathcal{C},\ \forall t \in \mathcal{T},\ \forall \delta \in \Delta:
\]
\begin{multline}
\textstyle
x^{v,c, \delta}_{t} \leq a^{v,c}_{t}; \quad
x^{v,c, \delta}_{t} \leq y^{v, \delta}_t; \quad
x^{v,c, \delta}_{t} \geq a^{v,c}_{t} + y^{v, \delta}_t - 1
\end{multline}
\begin{equation}\label{eq:c3}
\textstyle
\forall v \in \mathcal{V},\ \forall t \in \mathcal{T}^v: 
r^v_t \leq \eta \sum_{c \in \mathcal{C}} \sum_{\delta \in \Delta} r^{v,\delta}_{\max} \cdot x^{v,c, \delta}_{t}
\end{equation}
where \( \eta \) denotes the round-trip efficiency of charging and discharging, and \( r^{v, \delta}_{\max} \) represents the maximum charging rate for EV \( v \) within SoC segment \( \delta \), expressed as a fraction of the charger’s capacity \( c_{\max} \).

The minimum charging rate is set to be the minimum rate supported by the connected charger:
\begin{equation}\label{eq:c4}
\forall v \in \mathcal{V},\ \forall t \in \mathcal{T}^v: \quad
r^v_t \geq \eta \sum_{c \in \mathcal{C}} c_{\min} \cdot a_{v,c,t}
\end{equation}

\noindent The SoC update over time follows:
\begin{equation}\label{eq:e_s0_new}
\textstyle
\forall v \in \mathcal{V},\ \forall t \in \mathcal{T}^v: \quad
e^v_{t_n} =
\begin{cases}
e^v_{t_{arr}} + r^v_{t_n}, & \text{if } n = 0 \\[8pt]
e^v_{t_{n-1}} + r^v_{t_n}, & \text{otherwise}
\end{cases}
\end{equation}
where, $e_{0}^{v}$ is the initial charge the EV arrives with.

To discourage excessive discharging, the total discharge for each EV is tracked using:
\begin{equation}\label{eq:disch_track}
q^v_t \geq -r^v_t
\end{equation}

\noindent\emph{Electricity Cost and SoC Completion. }
The energy cost to the building during each time step combines EV charging loads and the building's base load under the time-of-use price:
\begin{equation}\label{eq:energy_cost}
\textstyle
\forall t \in \mathcal{T}: \quad
g_t = \sum_{v \in \mathcal{V}} w_t \left( c^v_t + {p^b_t}/{\tau} \right)
\end{equation}

To ensure user satisfaction, each EV must reach its desired SoC before departure, and deviations are penalized:
\begin{equation}\label{eq:soc_req}
\begin{split}
\noindent
\forall v \in \mathcal{V}: \quad
z_v =  \left| e^v_{\text{req}} - e^v_{t_{\text{dep}}} \right|\\
\implies z_v \geq e^v_{\mathrm{req}} - e^v_{t_{\mathrm{dep}}}; \quad
z_v \geq e^v_{t_{\mathrm{dep}}} - e^v_{\mathrm{req}} \\
\end{split}
\end{equation}

\noindent\emph{Demand Cost Calculation.} 
To compute the demand cost, we track the power use of the building and chargers combined.
The demand cost is calculated on the maximum power.
\begin{equation}\label{eq:pi_s}
\textstyle
\forall t \in \mathcal{T}: \quad
p_t = p^b_t + \frac{1}{\tau} \sum_{v \in \mathcal{V}} r^v_t
\end{equation}
\begin{equation}\label{eq:c9}
\textstyle
\forall t \in \mathcal{T}: \quad
p^{\max} \geq p_t
\end{equation}

\noindent{\bf Objective Function. } 
The overall objective is to minimize the total electricity cost (energy cost + demand cost) and penalties for missed SoC targets:
\begin{equation}\label{eq:obj2}
\begin{split}
\textstyle
\min \sum_{t\in \mathcal{T}} g_t + K_{\mathrm{DC}} \cdot p^{\max} + K_{\mathrm{SOC}} \cdot \sum_{v \in \mathcal{V}} z_v \\
+ K_{batt} \sum_{v \in \mathcal{V}} \sum_{t \in \mathcal{T}^v} q^v_t
\end{split}
\end{equation}
where \( K_{\mathrm{DC}} \) is the demand charge rate, \( K_{\mathrm{SOC}} \) is the compensation for unfulfilled SoC (per kWh) and controls the penalty for unfulfilled SoC targets. $K_{\mathrm{batt}}$ is the penalty rate per kWh discharged to limit excessive battery wear.

\subsection{Online Control}

In real-world deployment, we consider the problem of optimizing EV charging decisions in a dynamic environment, where the static optimization method described earlier becomes inapplicable due to the absence of oracle inputs. To handle this online decision-making process under uncertainty, we propose a Monte Carlo Model Predictive Control (MC-MPC) framework, which belongs to the family of stochastic Model Predictive Control (SMPC) methods~\cite{MA2016}. This framework extends our static optimization approach to an online setting by leveraging predictive models to simulate future scenarios.

\noindent{\bf Online EV Charging Optimization.} 
To support real-time operation, we reformulate the offline MILP problem as a sequential decision-making process, where the system evolves over discrete time steps. At each time step $t$, the system is characterized by the current state: 
$
S_t =(
\{e^v_t,\; e^v_{\text{req}},\; e^v_{\min},\; e^v_{\max},\; t^v_{\text{dep}},\; \phi_t(v) \}_{v \in \mathcal{V}_t},\;
b^p_t,\;
{p}_\text{past}^{\max})
$ where, $V_t \subseteq \mathcal{V}$ denotes the set of EVs present by time $t$, and for each EV $v \in \mathcal{V}_t$:  
$e^v_t$ is the current state-of-charge (SoC),  
$e^v_{\text{req}}$ is the requested SoC at departure,  
$e^v_{\min}$ and $e^v_{\max}$ are the SoC bounds,  
$t^v_{\text{dep}}$ is the scheduled departure time,  
$\phi_t(v)$ indicates the charger assigned to $v$, and 
The building state includes the current energy usage $b^p_t$ and the past peak demand ${p}_\text{past}^{\text{max}}$.

Based on the current state, the policy selects an action:
$ A_t = \left( r^v_t \;\middle|\; v \in \mathcal{V}_t \right)$, 
where $r^v_t$ is the assigned charging or discharging energy for each EV $v$ present at time $t$. 
The system then transitions to the next state $s_{t+1}$ based on the applied charging actions, EV arrivals and departures, battery dynamics (as defined in Eqs.~(\ref{eq:car_charger})-~(\ref{eq:c9})) and the updated building load. 

\noindent{\bf MC-MPC Controller.} We define the MC-MPC controller as a policy function
$\pi_{\text{ch}}: (S_t, \mathcal{F}) \rightarrow A_t$,
where $S_t$ is the current system state and $\mathcal{F}$ is a set of Monte Carlo-sampled future trajectories. Given these inputs, the controller selects a feasible action $A_t$ that minimizes the expected cumulative cost over the remainder of the billing period, based on outcomes simulated from $\mathcal{F}$. 
Each trajectory $f \in \mathcal{F}$ simulates the system’s evolution over a prediction horizon $[t, H(t)]$, where $H(t)$ denotes the final time step of the current decision period (the end of the day for our problem). The simulation includes future EV arrivals, their departure times and SoC target, as well as building load profiles, using the generation models (described in Section~\ref{sec:data_gen}). Let $\mathcal{V}^f = \mathcal{V}_t \cup \mathcal{V}_{t+1:H(t)}^f$ represent the set of all EVs (existing and future) in trajectory $f$.

At each time step, we aim to determine the current action $A_t$ by solving a short-term static optimization problem, which, given the current EV status specified in $S_t$, minimizes the expected cost across all sampled future trajectories $\mathcal{F}$, coupled at the current state. While each trajectory $f$ may follow its own optimal actions from $t+1$ to $H(t)$, the action at the current time step $A_t$ must remain consistent across all trajectories. This leads to the following optimization formulation:

\begin{align}
\label{eq:mpc_objective}
\min_{\substack{A_t,\, \{A_{t+1}^f, \dots, A_{H(t)}^f\}_{f \in \mathcal{F}}}} 
\; \mathbb{E}_{f \in \mathcal{F}} \Bigg[
& \sum_{k = t}^{H(t)} g_{k,f} 
+ w^d p^{\max}_f \\
+ K_{\mathrm{SOC}} \sum_{v \in \mathcal{V}^f} z_v^f \nonumber
& + K_{\mathrm{batt}} \sum_{v \in \mathcal{V}} \sum_{t \in \mathcal{T}^v} q^v_{t,f}
\Bigg]
\end{align}

subject to the EV-charger assignment feasibility constraints.
Here, $A_t$ is the common action applied at time $t$, while $\{{A_{t+1}^f, \dots, A_{H(t)}^f}\}$ are trajectory-specific future actions. The term $g_k^f$ denotes the energy cost at time step $k$ in trajectory $f$, $p^{\max}_f$ is the peak demand over the trajectory (including the historical peak $p^{\max}_{\text{past}}$), and $z_v^f$ represents the SoC shortfall for EV $v$ upon its departure in trajectory $f$.

\begin{theorem} {\bf Temporal decomposition}. Minimizing daily peak demand independently each day leads to a minimized monthly demand cost, without requiring multi-day coupling. 
\begin{proof} 
Our MC-MPC approach minimizes the monthly total cost by decomposing the problem into daily control tasks and optimizing each day independently. This is supported by (i) the system assumption that EVs do not persist across days in the building, and (ii) the property of the demand charge structure that the monthly peak can occur on any day, allowing each day’s peak to contribute independently to the monthly maximum.\\
Let \(\mathcal{D}\) denote the set of days in the monthly billing cycle, and for each \(d\in\mathcal{D}\), let \(\mathcal{T}_d\) denote the time steps belonging to day \(d\).
Define the full monthly time horizon as \(\mathcal{T}^{\text{month}} = \bigcup_{d \in \mathcal{D}} \mathcal{T}^{d}\). 
The monthly peak demand is given by: $ P_{\max}^{\text{month}} = \max_{t \in \mathcal{T}^{\text{month}}} (\pi_t)$, where \(\pi_t\) stores the instantaneous total power use at time \(t\).
By properties of the maximum operator over disjoint sets, we have:
\[
\textstyle
\max_{t \in \mathcal{T}^{\text{month}}} \pi_t = \max_{t \in \bigcup_{d\in\mathcal{D}} \mathcal{T}^{(d)}} \pi_t = \max_{d \in \mathcal{D}} \left( \max_{t \in \mathcal{T}^{d}} \pi_t \right)
\]
We can defining the daily peak load as $P_{\max}^{d} = \max_{t \in \mathcal{T}^{d}} (\pi_t)$
and it follows:
$$
\textstyle
P_{\max}^{\text{month}} = \max_{d \in \mathcal{D}} P_{\max}^{d}.
$$ 
\noindent We formally show that minimizing daily peak loads independently \(P_{\max}^{d}\) minimizes the overall monthly peak \(P_{\max}^{\text{month}}\).  
\end{proof}

\end{theorem}

\noindent{\bf Action Refinement.} To further reduce the monthly demand cost while meeting EV SoC requirements, we introduce an action refinement step defined as a function $f_{\text{refine}} : (A_t, \hat{P}_{\max}) \rightarrow \tilde{A}_t$, which adjusts the action $A_t$ based on the estimated peak power $\hat{P}_{\max}$. If the current aggregate power (building load plus charging) is below $\hat{P}_{\max}$, charging rates can be safely increased without incurring additional demand costs. This proactive adjustment helps fulfill SoC targets, discharge at opportune times and reduce the likelihood of higher peaks later in the billing period as EVs approach departure.  
Specifically, we estimate the gap between the predicted peak power and the current total charging rate: $P_{\text{diff}} = \hat{P}_{\max} - (\sum_{v \in \mathcal{V}_t} r_t^v /\tau + b^p_t) $. If $P_{\text{diff}} > 0$, we distribute this remaining capacity equally across all connected EVs while respecting their charging constraints and SoC limits. The final charging rate $\tilde{A}_t$ for each EV $v \in \mathcal{V}_t$ is adjusted by: 
\begin{equation} 
\tilde{r}_t^v =
\begin{cases}
r_t^v +  \min \left(\frac{P_{\text{diff}}}{|\mathcal{V}_t|} \tau,\ e_{\text{req}}^v,\ \phi(v)_{{max}}\right), & \text{if } \phi(v)_{{min}} = 0 \ \\
r_t^v + \min \left(\frac{P_{\text{diff}}}{|\mathcal{V}_t|} \tau,\ e_{\max}^v,\ \phi(v)_{\max} \right), & \text{if } \phi(v)_{{min}} < 0\
\end{cases} 
\label{eq:refine}
\end{equation}
for $v\in \mathcal{V}_t$ where $\phi(v) = 0$ denotes a uni-directional charger, and $\phi(v) < 0$ denotes a bi-directional charger. 
Algorithm~\ref{alg:MP-MPC} provides an overview of our V2B charging optimization framework. 

\begin{algorithm}[ht]
\caption{Online V2B Charging Optimization}
\label{alg:MP-MPC}
\KwIn{Initial state $S_0$; number of samples $N$; billing period $T$; estimated monthly peak power $\hat{P}_{\max}$}
\KwOut{Charging decisions $\{A_0, A_1, \dots, A_T\}$}

\For{$t = 0$ \KwTo $T$}{
    Check EV departures and free assigned chargers\;
    Assign chargers to new arrivals using FIFO; prioritize bidirectional chargers\;
    Observe current state $S_t$\;
    Future trajectory set $\mathcal{F} \leftarrow \emptyset$\;
    
    \For{$i = 1$ \KwTo $N$}{
        Sample trajectory $f$ from the generative model given $S_t$\;
        Simulate from $t{+}1$ to $H(t)$ and add $f$ to $\mathcal{F}$\;
    }
    
    Get $A_t$ by solving the optimization in Eq.~(\ref{eq:mpc_objective})\;
    Refine action to $\tilde{A}_t = f_{\text{refine}}(A_t, \hat{P}_{\max})$ by Eq.~(\ref{eq:refine})\;
    Apply $\tilde{A}_t$ and update system state to $S_{t+1}$\;
}
\end{algorithm}

\section{Optimization Framework for Negotiation}\label{sec:neg_milp}

In this section, we present the optimization framework that underpins our negotiation-based Model Predictive Control (MPC) approach for managing and providing options in the V2B electric vehicle charging environment. The framework formulates the optimization problem as a Mixed-Integer Linear Program (MILP) that jointly considers user charging preferences (deviation limits), building operational constraints, and dynamic electricity pricing and demand charges.

The following subsections detail the MILP formulation, scenario sampling procedures, and the solution methodology that iteratively adapts charging and incentive offers at each decision epoch.

\noindent\textbf{Decision Variables} 

We define the following decision variables, distinguishing between those reused from the charging policy and those introduced for the negotiation policy. We reuse some variables from the charging policy, and define more new variable specific to the negotiations case. Variables inherited from the charging policy are:
\begin{itemize}
    \item $a^{v,c}_{t} \in \{0,1\}$: Indicates whether electric vehicle (EV) $v$ is assigned to charger $c$ at time step $t$.
    \item $r^v_t \geq 0$: Charging power assigned to EV $v$ at time step $t$.
    \item $e^v_{t} \geq 0$: State of charge (SoC) of EV $v$ at time step $t$.
    \item $t_{\mathrm{dep}}^{v}$: Optimized departure time for EV $v$.
    \item $e_{\mathrm{dep}}^{v} \geq 0$: SoC of EV $v$ at its (possibly negotiated) departure time.
\end{itemize}

In the negotiation problem, we require additional variables since the departure time and requested state of charge (SoC) are not fixed. Instead, for each option $l \in L$, we know only the allowable deviation limits. These limits are used to identify the optimal combination of user utility, departure time, and SoC.
\begin{itemize}
    \item $\lambda^v_{t} \in \{0,1\}$: Indicates whether EV $v$ departs at time step $t$.
    \item $\sigma^v_{soc} \in [0, e^{\mathrm{max}}_l]$: Deviation of SoC at departure from the required SoC for EV $v$, bounded by limit $l$.
    \item $\sigma^v_{dep} \in [0, t^{\mathrm{max}}_{\mathrm{dep}, l}]$: Deviation of actual departure time from the required departure time for EV $v$, bounded by limit $l$.
\end{itemize}

\noindent\textbf{Constraints}

\begin{enumerate}

\item As discussed earlier, at each time step $t$, chargers and EVs are matched one-to-one: each charger serves at most one EV, and each EV connects to at most one charger based on the first-come-first-served (FCFS) assignment policy. Formally, this defines the mapping $\phi_t : \mathcal{V}_t \leftrightarrow \mathcal{C}_t$,

\item Similar to equation~\eqref{eq:car_charger}, the charger assigned to EV $v$ must remain the same throughout its duration of stay. However, since the exact departure time is unknown, we introduce $\lambda$ to determine the optimal departure time:

\begin{align}
\forall v \in \mathcal{V},\ \forall c \in \mathcal{C},\ \forall t \in \mathcal{T}^v , t > t^v_{\text{arr}}:\\
\quad a^{v, cp}_{t-1} \leq a^{v, cp}_{t} + \lambda^v_{t}
\end{align}

\item {To prevent assignments after departure}:
\begin{equation}
\forall v \in \mathcal{V},\ \forall c \in \mathcal{C},\ \forall t \in \mathcal{T}^v: \quad a_{v, c, t} \leq 1 - \sum_{t=t^v_{\text{arr}}}^{t} \lambda^v_{t}
\end{equation}

\item Charging is done only when assigned, and follows the exact constraints from the charging policy. Thus, we apply equations~\eqref{eq:soc_seg} - \eqref{eq:e_s0_new} to maintain the charging and discharging rates in each SoC segment of the piece-wise linear SoC curve.

\item Any discharge accounts towards battery wear and is controlled similar to the charging policy, using equation~\eqref{eq:disch_track}.

\item {Since the problem decides the optimized departure time step within the duration of stay, it is tracked using}:
\begin{equation}
\forall v \in \mathcal{V}: \quad t_{\text{dep}}^{v} = \sum_{t \in \mathcal{T}^v} t \cdot \lambda_{v,t}
\end{equation}

\item {We also need to track and make sure that each vehicle departs exactly once, within the maximum limit of its duration $\mathcal{T}^v$}:
\begin{equation}
\forall v \in \mathcal{V}: \quad \sum_{s \in \mathcal{T}^v} \lambda^v_{t} = 1
\end{equation}

\item Within the allowed deviation limits, we need to find the optimal deviations in SoC and departure time. The following equations keep track of these deviations:
 
\begin{equation}
\forall v \in \mathcal{V}: \quad \sigma_{soc}^v =  e_{\text{req}}^v - e_{\text{dep}}^v
\end{equation}

\begin{equation}
\forall v \in \mathcal{V}: \quad \sigma_{dep}^v = t_{\text{dep}}^v - t_{\text{req}}^v
\end{equation}

\item Energy cost calculation for the building:

We follow the same constraints in equations~\eqref{eq:energy_cost} - \eqref{eq:soc_req} to calculate the energy cost.

\item Demand cost calculation for the building:

Equations~\eqref{eq:pi_s} - \eqref{eq:c9} are used to calculate the demand cost.
\end{enumerate}

\noindent{\bf Objective Function. } 
The overall objective is again the same as in the charging optimization, i.e., to minimize the total cost cost in eq.~\eqref{eq:obj}, to minimize energy, demand cost, penalties for missed SoC targets, and excessive discharge:

\begin{equation}\label{eq:obj3}
\begin{split}
\textstyle
\min \sum_{t\in \mathcal{T}} g_t + K_{\mathrm{DC}} \cdot p_{\max} + K_{\mathrm{SOC}} \cdot \sum_{v \in \mathcal{V}} z_v \\
+ K_{batt} \sum_{v \in \mathcal{V}} \sum_{t \in \mathcal{T}^v} q^v_t
\end{split}
\end{equation}
where \( K_{\mathrm{DC}} \) is the demand charge rate, \( K_{\mathrm{SOC}} \) is the compensation for unfulfilled SoC (per kWh) and controls the penalty for unfulfilled SoC requirements. $K_{\mathrm{batt}}$ provides the battery degradation penalty for discharging.

\textbf{Practical Considerations:}  
This approach does not require perfect foresight; it only relies on the latest forecasts, updating its policy as new data arrive.

At every step, MPC with Monte Carlo sampling ensures that the negotiation process is both flexible and tractable. The system computes user-specific incentives and charging schedules that align with both real-time operational requirements and user behavior, all while maintaining scalability to large fleets and adaptability to new data or planning advancements.

\color{black}

\section{Proof of Theorems}\label{sec:proofs}

\textsc{Strategy-Proofness for Departure Time and SoC}

\emph{The mechanism is strategy-proof with respect to departure time and requested SoC: a user cannot increase their utility by misreporting an earlier departure time or a higher required SoC \(\bar{\theta}_v\) than their true preferences \(\theta_v\). Formally, truthful reporting always maximizes user utility, i.e., \(U(\theta_v) \geq U(\bar{\theta}_v)\).}

\begin{proof}[Proof]
Consider a user $v$ whose utility is defined as the marginal reduction in system cost due to their participation:
$$
U(\theta_v) = J^{\pi^*}(\boldsymbol{\theta}_{\setminus v}, \boldsymbol{\hat{\theta}}_{\setminus v}) - J^{\pi^*}(\boldsymbol{\theta}_{\setminus v}, \theta_v, \boldsymbol{\hat{\theta}}_{\setminus v}).
$$
Here $J^{\pi^*}(\boldsymbol{\theta}_{\setminus v}, \boldsymbol{\hat{\theta}}_{\setminus v})$ denotes the expected cost of the system without user $v$ (but including the other users), and $J^{\pi^*}(\boldsymbol{\theta}_{\setminus v}, \theta_v, \boldsymbol{\hat{\theta}}_{\setminus v})$ the cost with user $v$ participating truthfully.

\textbf{Departure Time.}
Suppose user $v$ reports a departure time $\bar{t}_{\text{dep}} \in \bar\theta_v$ that is earlier than their true departure time $t_{\text{dep}} \in \theta_v$ ($\bar{t}_{\text{dep}} < t_{\text{dep}}$).
This act constrains the set of possible charging actions for the system because the charging schedule for $v$ must now conclude by $\bar{t}_{\text{dep}}$ rather than $t_{\text{dep}}$. 
Formally, the set of time steps available for charging is reduced from the interval $[t_{\mathrm{arr}}, t_{\text{dep}}]$ to $[t_{\mathrm{arr}}, \bar{t}_{\text{dep}}]$.

Accordingly, the optimization is conducted over a reduced set of time steps for user $v$, thereby limiting the system's temporal flexibility to leverage lower-priced periods or to achieve better peak load flattening. Any load shifting or peak smoothing achievable within the shortened horizon $[t_{\mathrm{arr}}, \bar{t}_{\mathrm{dep}}]$ is also feasible over the extended horizon $[t_{\mathrm{arr}}, t_{\mathrm{dep}}]$. Thus, allowing the user to remain until $t_{\mathrm{dep}}$ (with $t_{\mathrm{dep}} > \bar{t}_{\mathrm{dep}}$) provides additional opportunities for load shifting and peak flattening, potentially reducing overall costs. Since the system always minimizes the total cost within the available set of charging actions, restricting the window cannot lower this minimum cost. Thus,

$$
J^\pi(\theta_{\setminus v}, \bar{\theta}_v, \boldsymbol{\hat{\theta}}_{\setminus v}) \geq J^\pi(\theta_{\setminus v}, \theta_v, \boldsymbol{\hat{\theta}}_{\setminus v}) \text{ and,}
$$
\begin{align}\label{eq:proof1_1}
U(\bar \theta_v) = J^{\pi^*}(\boldsymbol{\theta}_{\setminus v}, \boldsymbol{\hat{\theta}}_{\setminus v}) - J^{\pi^*}(\boldsymbol{\theta}_{\setminus v}, \bar \theta_v, \boldsymbol{\hat{\theta}}_{\setminus v}).
\end{align}

Subtracting the utilities,
\begin{align}\label{eq:proof1_2}
U(\theta_v) - U(\bar{\theta}_v) = J^\pi(\theta_{\setminus v}, \bar{\theta}_v, \boldsymbol{\hat{\theta}}_{\setminus v}) - J^\pi(\theta_{\setminus v}, \theta_v, \boldsymbol{\hat{\theta}}_{\setminus v}).
\end{align}

Therefore,
\begin{align}\label{eq:proof1_3}
U(\theta_v) - U(\bar{\theta}_v) \geq 0,
\end{align}

which implies
\begin{align}\label{eq:proof1_4}
U(\theta_v) \geq U(\bar{\theta}_v).
\end{align}

Thus, the utility for misreporting is less than or equal to that for reporting truthfully. We prove that reporting the true departure time maximizes user utility.

\textbf{State of Charge (SoC).}
Similarly, if user $v$ reports a higher required SoC $\bar{e}^v_{\mathrm{req}} \in \bar{\theta}_v$ at departure than their true requirement $e^v_{\mathrm{req}} \in \theta_v$, i.e., ($\bar{e}^v_{\mathrm{req}} > e^v_{\mathrm{req}}$), the system must deliver a greater amount of energy within the same (or possibly even a reduced) time window. This imposes a stricter set of feasible charging schedules, as the minimum total energy to be supplied increases. The increased constraint limits the system's ability to optimize cost, and so
$$
J^\pi(\theta_{\setminus v}, \bar{\theta}_v, \boldsymbol{\hat{\theta}}_{\setminus v}) \geq J^\pi(\theta_{\setminus v}, \theta_v, \boldsymbol{\hat{\theta}}_{\setminus v}) \text{ and,}
$$
Hence, similar to the previous proof, by following steps \eqref{eq:proof1_1}-\eqref{eq:proof1_3},  we can prove that the user cannot achieve greater utility by exaggerating their SoC requirement, and, 
\[
U(\theta_v) \geq U(\bar{\theta}_v).
\]

\textbf{Calculation of Utility Difference.}
Combining both cases, for any misreported parameter $\bar{\theta}_v$, the utility difference obeys
\[
U(\theta_v) - U(\bar{\theta}_v) \geq 0.
\]
Thus, no user benefits by tightening their reported constraints; truthful reporting of departure time and SoC is optimal.

\end{proof}

\textsc{Bounding Utility by Total Savings}

\emph{Any payment $\zeta^v$ to user $v \in \mathcal{V}$ must satisfy $\zeta^v \ge -U(\theta_v)$ to ensure that the building’s total cost after incentives does not exceed the baseline.}

\begin{proof}
    
We begin from the {budget feasibility constraint}, which requires that the building's cost with user $v$ and payment $\zeta^v$ does not exceed the cost without $v$:
\begin{equation}
    J^{\pi^*}(\boldsymbol{\theta}_{\setminus v}, \theta_v, \boldsymbol{\hat{\theta}}_{\setminus v}) - \zeta^v \le J^{\pi^*}(\boldsymbol{\theta}_{\setminus v}, \boldsymbol{\hat{\theta}}_{\setminus v}).
\end{equation}
Rearranging terms:
\begin{equation}
    -\zeta^v \le J^{\pi^*}(\boldsymbol{\theta}_{\setminus v}, \boldsymbol{\hat{\theta}}_{\setminus v}) - J^{\pi^*}(\boldsymbol{\theta}_{\setminus v}, \theta_v, \boldsymbol{\hat{\theta}}_{\setminus v}) = U(\theta_v).
\end{equation}
Thus, we obtain the bound:
\begin{equation}
    \zeta^v \ge -U(\theta_v).
\end{equation}
which is valid through Eq.~\eqref{eq:user_cost}.

This ensures that the payment to user $v$ does not exceed the cost savings they bring. In the limiting case (equality), we recover the identity:
\begin{equation}
    J^{\pi^*}(\boldsymbol{\theta}_{\setminus v}, \theta_v, \boldsymbol{\hat{\theta}}_{\setminus v}) + U(\theta_v) = J^{\pi^*}(\boldsymbol{\theta}_{\setminus v}, \boldsymbol{\hat{\theta}}_{\setminus v}).
\end{equation}

\end{proof}

\textsc{Voluntary Participation}

\emph{For all rational users $v$, the user choice mechanism guarantees ${Y}^v_{l} \ge 0$. Thus, no user is worse off by participating, and some benefit strictly if $U(\theta_v) > 0$.}

\begin{proof}

If $U(\theta_v)\geq0$, then from Eq.~\ref{eq:user_cost}, the user cost for option $l$ is less than the charging cost, $\zeta^v_l \leq \sum_{t\in \mathcal{T}^v} g^v_t w_t$. This ensures the user has a charging price lower than the time-of-use energy cost. 
\noindent

Also, recall the user satisfaction metric
\(Y^{n}_{l} = \bar E - (\zeta_{l}+I^{\,n}_{l})\)
from eq.~\ref{eq:user_satisfaction}.  Whenever
\[
\zeta_{l}+I^{\,n}_{l} \;>\; \bar E
\;\;\Longleftrightarrow\;\;
Y^{n}_{l}<0 ,
\]
the charging options at the building are strictly more expensive than the external alternative.  A rational user therefore rejects the contract and charges externally at cost \(\bar E\).  Consequently, the maximum amount a user can ever pay at the building is \(\bar E\), and participation in the building’s negotiation mechanism can never leave them worse off.

\end{proof}

\section{User Segmentation from Survey}\label{sec:survey_analysis}

\subsection{Flexibility limits for $L$ Negotiation Options}

To identify distinct user groups based on their flexibility preferences, we conducted a survey asking respondents about two key behavioral dimensions: their maximum acceptable departure delay and their maximum acceptable reduction in state-of-charge (SoC), or range flexibility, when offered incentives. These questions were designed to capture users' willingness to adjust their charging behavior under different conditions. All the survey questions are shared below.

\noindent\textbf{Clustering Methodology}

After collecting the survey responses, we preprocessed the data for clustering analysis. Responses for maximum departure delay, initially provided as ordinal categories from 1 to 9, were converted to corresponding time values in minutes (0, 15, 30, 45, 60, 75, 90, 105, 120 minutes). Similarly, range flexibility responses, originally one-hot encoded, were transformed into percentage values (0\%, 5\%, 10\%, 20\%) representing the maximum acceptable SoC reduction.

Data preprocessing also involved removing incomplete responses, such as those where users selected ``I need more info'' for range flexibility or provided no response. These could not be meaningfully incorporated into the distance-based clustering algorithm. This filtering resulted in a final dataset of 27 valid responses across both dimensions.

Given the different scales of the two variables (minutes versus percentages), we applied standardization using z-score normalization to ensure equal weighting in the clustering algorithm. This step is critical for K-means clustering, as the algorithm is sensitive to variable scales and may otherwise be dominated by the variable with larger numerical range.

Finally, we applied K-means clustering to the standardized data to segment users into homogeneous groups. These clusters provide valuable insights for developing targeted charging strategies tailored to the flexibility profiles of different user segments.

\noindent\textbf{Optimal Cluster Determination}

To identify the optimal number of clusters, we applied two widely-used validation techniques: the \emph{silhouette method} and the \emph{elbow method}. Both approaches help assess cluster quality by analyzing solutions over a range of cluster counts ($k = 2$ to $k = 7$).

The \textit{silhouette score} quantitatively measures how well each data point fits within its assigned cluster. For a data point $i$, its silhouette value $s(i)$ is given by:
\[
s(i) = \frac{b(i) - a(i)}{\max\{a(i),\, b(i)\}}
\]
where $a(i)$ is the average distance from $i$ to all other points in its own cluster, and $b(i)$ is the lowest average distance from $i$ to points in any other cluster (i.e., the nearest neighboring cluster). The silhouette score ranges from $-1$ to $1$, where a higher value indicates better-defined and more cohesive clusters.

The \textit{elbow method} involves plotting the total within-cluster sum of squared distances (WCSS) for different values of $k$. As $k$ increases, WCSS decreases because clusters contain fewer points and become more compact. However, after a certain point, the rate of decrease sharply slows, forming an ``elbow'' which suggests that additional clusters beyond this point yield limited improvement and may not be meaningful.

\vspace{0.5em}
\begin{center}
\footnotesize
\label{tab:silhouette_scores}
\begin{tabular}{c|c}
\toprule
\textbf{Number of Clusters ($k$)} & \textbf{Silhouette Score} \\
\midrule
2 & 0.498 \\
3 & 0.588 \\
4 & 0.547 \\
5 & 0.563 \\
6 & 0.589 \\
7 & 0.596 \\
\bottomrule
\end{tabular}
\end{center}
\vspace{0.5em}

In our analysis (Figure~\ref{fig:silhouette}), the highest silhouette score was observed at $k=7$, suggesting that this value offers the most well-defined partitioning. However, $k=3$ yields a comparable silhouette score and provides greater applicability to our problem. The elbow method also supports $k=3$ as a suitable choice (see Figure~\ref{fig:elbow_plot}). Therefore, we considered both mathematical criteria and domain interpretability when selecting the optimal number of clusters.

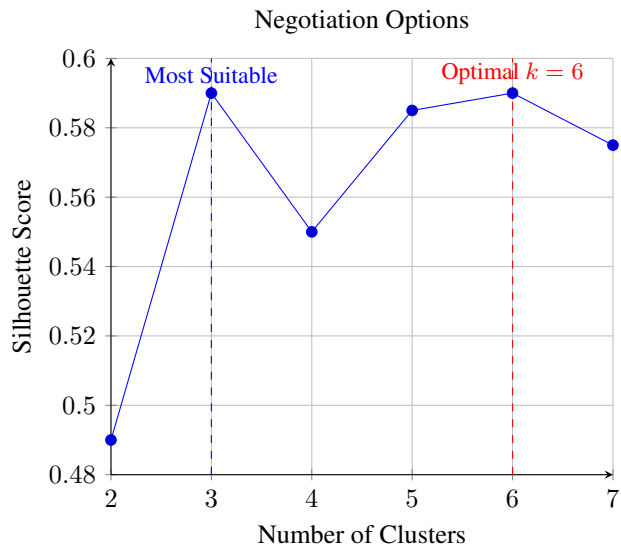
\begin{figure}[htbp]
\centering
\begin{tikzpicture}
\begin{axis}[
    width=0.5\columnwidth,
    xlabel={Number of Clusters},
    ylabel={Silhouette Score},
    title={Negotiation Options},
    grid=major,
    ymin=0.48, ymax=0.60,
    xtick={2,3,4,5,6,7},
    yticklabel style={/pgf/number format/fixed},
    axis x line=bottom,
    axis y line=left
]
\addplot+[mark=*] coordinates {
    (2,0.49) (3,0.59) (4,0.55) (5,0.585) (6,0.59) (7,0.575)
};

\addplot[dashed, red] coordinates {(6,0.48) (6,0.60)};
\node[red] at (axis cs:6,0.59) [above] {\footnotesize Optimal $k=6$};

\addplot[dashed, blue] coordinates {(3,0.48) (3,0.60)};
\node[blue] at (axis cs:3,0.59) [above] {\footnotesize Most Suitable};
\end{axis}
\end{tikzpicture}
\caption{Silhouette method shows $k=6$ as the optimal choice (highest score). However, $k=3$ is numerically very close, and is most suitable as users might get overwhelmed with more options.}
\label{fig:silhouette}
\end{figure}

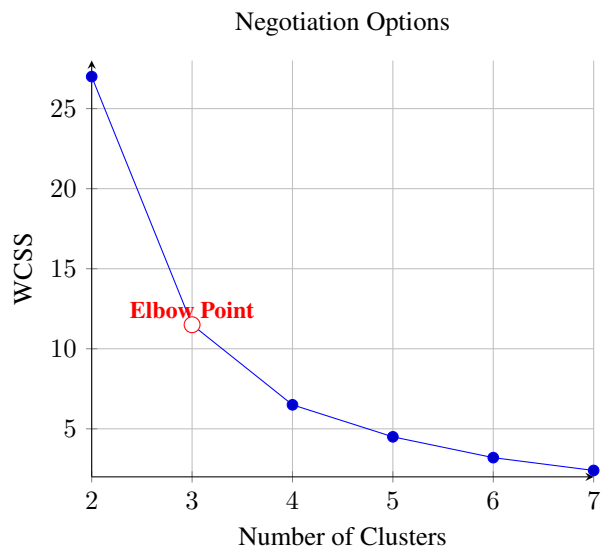
\begin{figure}[htbp]
\centering
\begin{tikzpicture}
\begin{axis}[
    width=0.5\columnwidth,
    xlabel={Number of Clusters},
    ylabel={WCSS},
    title={Negotiation Options},
    grid=major,
    ymin=2, ymax=28,
    xtick={2,3,4,5,6,7},
    axis x line=bottom,
    axis y line=left
]
\addplot+[mark=*] coordinates {
    (2,27) (3,11.5) (4,6.5) (5,4.5) (6,3.2) (7,2.4)
};

\addplot[
    only marks,
    mark=*, mark size=3pt,
    draw=red, fill=white
] coordinates {
    (3,11.5)
};
\node[red] at (axis cs:3,12.4) {\footnotesize \textbf{Elbow Point}};

\end{axis}
\end{tikzpicture}
\caption{Elbow method shows a sharp decrease until $k=3$, indicating the elbow point.}
\label{fig:elbow_plot}
\end{figure}

\begin{figure}[htbp]
    \centering
    \includegraphics[width=0.6\linewidth]{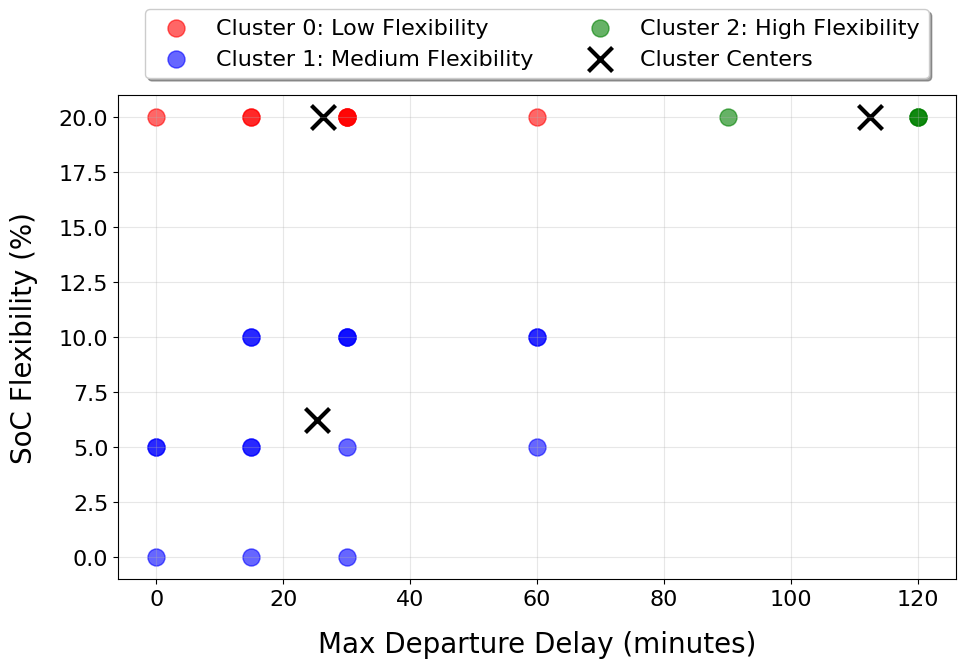}
    \caption{Clustered user deviation limits data provides Negotiation Options and flexibility limits}
    \label{fig:example_image}
\end{figure}

From a practical perspective, three clusters align well with common behavior groups in transportation research, representing low, medium, and high flexibility users. This approach offers useful insights for managing charging infrastructure without adding the complexity that more clusters would bring.

\begin{figure*}[htbp]
    \centering
    \includegraphics[width=\textwidth]{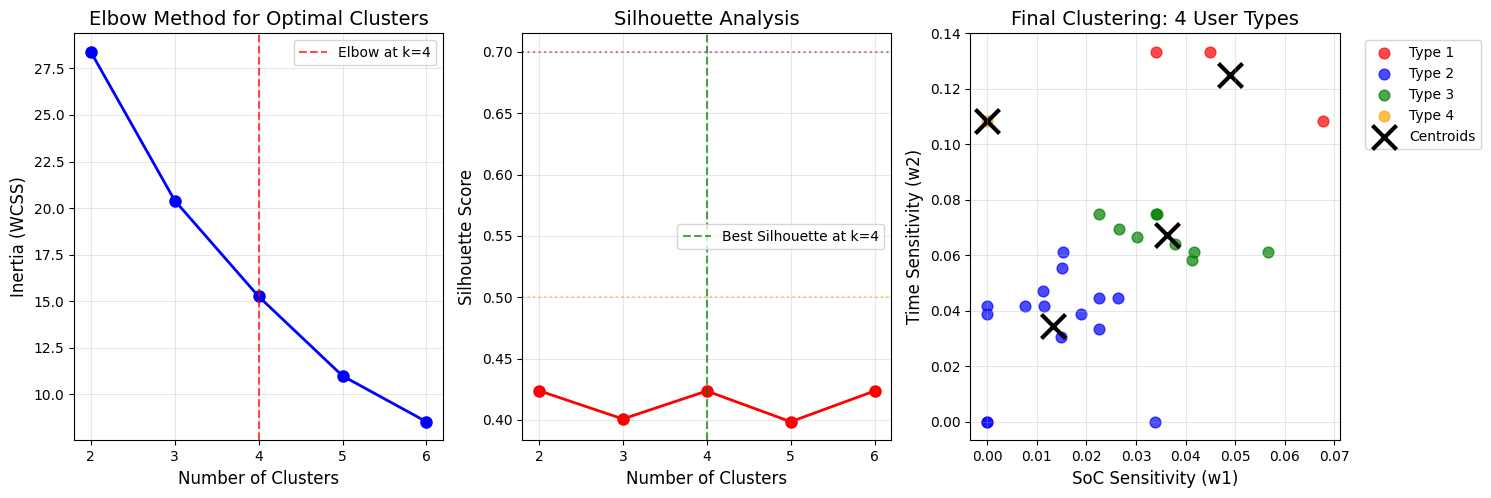}
    \caption{
Cluster selection and segmentation of user types. 
\textbf{Left:} The Elbow method plot shows inertia (WCSS) against the number of clusters, with an identified elbow at \(k = 4\).
\textbf{Center:} Silhouette scores for different cluster counts, indicating the best silhouette value at \(k = 4\).
\textbf{Right:} Final clustering of survey respondents into four user types based on SoC sensitivity (\(w_1\)) and time sensitivity (\(w_2\)), with cluster centroids marked by black crosses. Each color represents a distinct user type.
}
    \label{fig:user_type-clusters}
\end{figure*}

\subsection{Cluster Characterization and Representative Points}

The final three-cluster solution effectively segmented users into distinct behavioral profiles, as summarized in Table~\ref{tab:cluster_summary}, and is visualized in Figure~\ref{fig:user_type-clusters}. Each cluster was characterized by its centroid, the representative point that minimizes the within-cluster variance and serves as the archetypal user for that segment.

\begin{table}[h]
\centering
\footnotesize
\caption{Negotiation Options (Flexibility Analysis) Summary}
\label{tab:cluster_summary}
\begin{tabular}{c|c|cc|cc}
\toprule
\textbf{Cluster} & \textbf{Count} & \multicolumn{2}{c|}{\textbf{Dep. Flexibility (min)}} & \multicolumn{2}{c}{\textbf{SoC Flexibility (\%)}} \\
& & \textbf{Mean} & \textbf{Std} & \textbf{Mean} & \textbf{Std} \\
\midrule
1 & 7  & 27.86  & 18.22 & 20.00 & 0.00 \\
2 & 16 & 25.31  & 20.29 & 6.25  & 3.87 \\
3 & 4  & 112.50 & 15.00 & 20.00 & 0.00 \\
\bottomrule
\end{tabular}
\end{table}

\begin{table*}[htbp]
\centering
\begin{tabular}{l c c}
\toprule
\textbf{Scenario Type} & \textbf{Deviation Values} & \textbf{Example Base Cost (\$)} \\
\midrule
Departure delay (\(\Delta t_{\mathrm{dep}}\)) (minutes)    & 0, 15, 30, 60 & 10 \\
SoC reduction (\(\Delta e\)) (kWh)                         & 0, 5, 10, 20 & 6 -- 10 \\
\bottomrule
\end{tabular}
\caption{User Type Characteristics: Survey input points}
\label{tab:dev_scenarios}
\end{table*}

The cluster analysis revealed three distinct user segments:

\begin{itemize}
    \item \textbf{Cluster 1 - Low Flexibility Profile}: This segment (n=16, 59.3\% of sample) constitutes the largest user group, characterized by similar moderate departure delay tolerance (mean = 25.31 minutes) but substantially lower range flexibility (mean = 6.25\%). The higher standard deviation in departure delay (20.29 minutes) indicates greater heterogeneity within this cluster regarding time preferences, while the low standard deviation in range flexibility (3.87\%) suggests consistent preference for minimal SoC reduction.

    \item \textbf{Cluster 2 - Moderate Delay, High Range Flexibility}: This segment (n=7, 25.9\% of sample) represents users with moderate departure delay tolerance (mean = 27.86 minutes) but uniformly high range flexibility (20\%). These users demonstrate willingness to accept significant SoC reduction while maintaining relatively conservative time constraints.

    \item \textbf{Cluster 3 - High Delay, High Range Flexibility}: This smallest segment (n=4, 14.8\% of sample) exhibits the highest departure delay tolerance (mean = 112.50 minutes) combined with maximum range flexibility (20\%). The low standard deviation in departure delay (15.00 minutes) indicates homogeneous preferences for extended charging times within this group.
\end{itemize}

The cluster centroids capture typical user behaviors in each group. These centroids represent the ``typical'' user within each segment and can help guide charging infrastructure design and operation by enabling tailored services that match the needs of different user types.

\subsection{User Choice Modeling}

\noindent
\textbf{Building the Inconvenience Cost Model}

We use an anonymized stated-preference survey with 28 workplace EV users to quantify flexibility preferences and the monetary incentives required to accept various charging deviations. Each respondent indicated both how much they would need to save to tolerate additional departure delay and to accept reduced state-of-charge (SoC) at departure. The scenarios presented included:
\begin{itemize}
    \item \textbf{Departure delay} ($\Delta t_{\mathrm{dep}}$): Options for 15, 30, and 60 minutes, each with a base charging cost $\zeta = \$10$.
    \item \textbf{SoC reduction} ($\Delta e$): Reductions from an initial target of $80\%$ SoC to 75\% ($5\%$ reduction), 70\% ($10\%$), and 60\% ($20\%$), with scenario-specific base costs: $\zeta_{70\%} = \$8$, $\zeta_{75\%} = \$9$, $\zeta_{60\%} = \$6$.
\end{itemize}
Respondents mapped their categorical willingness to accept each deviation into corresponding discount percentages (ranging from 0\% for “no incentive needed” to 40\% for “extremely unlikely”), then to dollar amounts atop the relevant base cost. This processing yielded 168 valid scenario-user-incentive tuples $(\Delta e^{v,j},\ \Delta t^{v,j}_{\mathrm{dep}},\ I^{v,j})$. All the survey questions are shared below.

\medskip
\noindent
\textit{Step 1: Individual Inconvenience Cost Modeling}

For each participant $v$, we modeled their required incentive as a linear function of SoC reduction and time delay:
\[
I^{v,j} = w_1^{v}\, \Delta e^{v,j} + w_2^{v}\, \Delta t^{v,j}_{\mathrm{dep}} + \epsilon^{v,j}
\]
where $w_1^{v}$ (\$/\% SoC) and $w_2^{v}$ (\$/min) represent sensitivities to SoC and time deviations, and $\epsilon^{v,j}$ is a residual. We fit $(w_1^{v}, w_2^{v})$ for all 28 respondents with at least two usable scenarios using ordinary least squares, constraining the intercept to zero. This linear model explained a moderate share of variation in user responses (mean $R^2 = 0.57$; median $0.81$; quartiles: $0.36 / 0.81 / 0.90$), but also revealed that user heterogeneity is substantial.

\medskip
\noindent
\textit{Step 2: User Type Clustering and Summary}

To capture the diversity of user preferences at the population level, we standardized the fitted weights and applied K-means clustering to the survey respondents. After evaluating cluster solutions ranging from \(N=2\) to \(N=8\), we found that \(N=4\) clusters provided the best representation of distinct user types.
The optimal number of clusters was selected based on both the Elbow and Silhouette methods, with a WCSS of $15.25$ and a silhouette score of $0.424$. Both metrics indicate a fair level of cluster separation, suggesting reasonably distinct but not perfectly separated clusters.
 Figure~\ref{fig:user_type-clusters} displays the WCSS and silhouette scores, and the resulting clusters.
\begin{center}
\centering
\small
\label{tab:user_type_stats}
\begin{tabular}{c c c c}
\
{Type} & {SoC Sensitivity ($w_1$)} & {Time Sensitivity ($w_2$)} & {Share} \\
\midrule
i & $0.0489 \pm 0.02$ & $0.1250 \pm 0.01$ & 10.7\% \\
ii & $0.0133 \pm 0.01$ & $0.0346 \pm 0.02$ & 53.6\% \\
iii & $0.0362 \pm 0.01$ & $0.0673 \pm 0.01$ & 32.1\% \\
iv & $0.0000 \pm 0.00$       & $0.1083 \pm 0.00$         & 3.6\% \\
\hline
\end{tabular}
\end{center}

\noindent
These clusters reveal distinct user flexibility profiles:

\begin{itemize}
    \item {Type i} users are strongly averse to deviations in both departure time and SoC, representing highly inflexible behavior.
    \item {Type ii} users are the most flexible, showing the least sensitivity to changes in either parameter.
    \item {Type iii} users are moderately sensitive, willing to tolerate some deviation in both time and SoC, but not as much as Type II.
    \item {Type iv} users are a small minority who uniquely accept SoC deviations but remain inflexible regarding departure times.
\end{itemize}
This segmentation highlights clear differences in how users value their charging preferences and supports the development of more tailored management strategies.

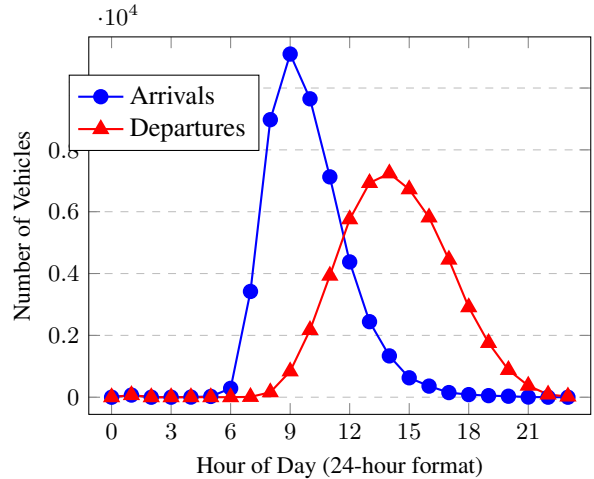
\begin{figure}[htbp]
\centering
\begin{tikzpicture}
\begin{axis}[
    width=0.5\textwidth,
    height=0.4\textwidth,
    xlabel={Hour of Day (24-hour format)},
    ylabel={Number of Vehicles},
    xtick={0,3,6,9,12,15,18,21},
    ymajorgrids=true,
    grid style=dashed,
    tick label style={font=\footnotesize},
    label style={font=\footnotesize},
    title style={font=\bfseries\large},
    legend style={at={(0.15,0.9)}, anchor=north, legend columns=1},
    legend cell align={left},
    enlargelimits=0.05,
]

\addplot[
    color=blue,
    mark=*,
    mark size=2.5pt,
    line width=0.8pt,
] coordinates {
    (0,0)    (1,70)   (2,0)    (3,1)    (4,4)    (5,25)
    (6,286)  (7,3421) (8,8976) (9,11099)(10,9651)(11,7127)
    (12,4376)(13,2443)(14,1336)(15,628) (16,359) (17,150)
    (18,84)  (19,51)  (20,32)  (21,6)   (22,1)   (23,0)
};
\addlegendentry{Arrivals}

\addplot[
    color=red,
    mark=triangle*,
    mark size=3pt,
    line width=0.8pt,
] coordinates {
    (0,0)     (1,69)   (2,0)    (3,0)    (4,1)    (5,1)
    (6,1)     (7,10)   (8,160)  (9,836)  (10,2173)(11,3931)
    (12,5755) (13,6929)(14,7240)(15,6725)(16,5815)(17,4450)
    (18,2910) (19,1755)(20,882) (21,373) (22,87)  (23,23)
};
\addlegendentry{Departures}

\end{axis}
\end{tikzpicture}
\caption{Hourly distribution of car arrivals and departures over all samples.}

\label{fig:car_arrival_departure}

\end{figure}

\begin{figure}[htbp]
\centering
\begin{tikzpicture}
\begin{axis}[
    width=0.5\textwidth,
    height=0.4\textwidth,
    xlabel={State of Charge (\%)},
    ylabel={Frequency},
    ymajorgrids=true,
    grid style=dashed,
    legend style={at={(0.4,0.9)}, anchor=north, legend columns=1},
    legend cell align={left},
    tick label style={font=\footnotesize},
    label style={font=\footnotesize},
    title style={font=\bfseries\large},
    enlargelimits=0.05,
    ymin=0,
    xmin=0,
    xmax=100,
]

\addplot+[
    ybar,
    bar width=3pt,
    bar shift=-2pt,
    fill=blue!30,
    draw=blue!80!black,
    area legend,
] coordinates {
    (2.5,  800)   (7.5, 1400)  (12.5,1900)  (17.5,2500)  (22.5,3100)
    (27.5,3600)  (32.5,4100)  (37.5,4600)  (42.5,5100)  (47.5,5400)
    (52.5,5800)  (57.5,6000)  (62.5,5800)  (67.5,5700)  (72.5,5400)
    (77.5,5000)  (82.5,4500)  (87.5,3900)  (92.5,3200)  (97.5,2500)
};
\addlegendentry{Initial SoC at Arrival}

\addplot+[
    ybar,
    bar width=3pt,
    bar shift=2pt,
    fill=red!20,
    draw=red!80!black,
    area legend,
] coordinates {
    (2.5,     0)   (7.5,     0)   (12.5,    0)   (17.5,    0)   (22.5,    0)
    (27.5,    0)   (32.5,    5)   (37.5,   20)   (42.5,  100)   (47.5,  400)
    (52.5, 2000)   (57.5, 4000)   (62.5, 7000)   (67.5,12000)   (72.5,20000)
    (77.5,35000)   (82.5,60000)   (87.5,75000)   (92.5,90000)   (97.5,95000)
};
\addlegendentry{Required SoC at Departure}

\addplot[blue, thick]  coordinates {(52.48,0)  (52.48,6500)};  \addlegendentry{Mean Arrival SoC: 52.48\%}
\addplot[red,  thick]  coordinates {(84.27,0)  (84.27,100000)};\addlegendentry{Mean Required SoC: 84.27\%}

\end{axis}
\end{tikzpicture}
\caption{Overlay of Initial SoC at Arrival and Required SoC at Departure (bins shifted for clarity).}
\label{fig:soc}

\end{figure}
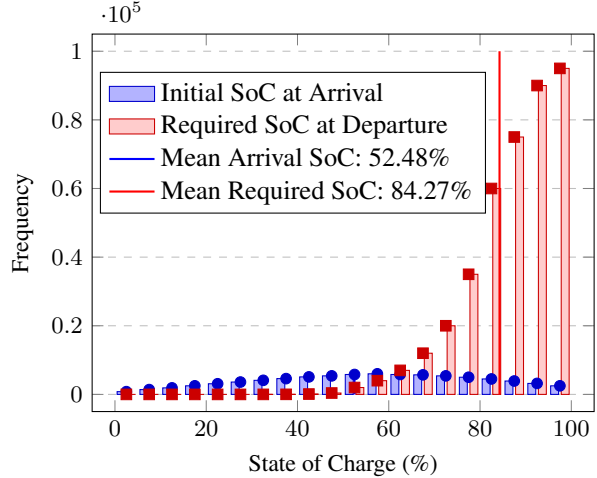

\section{Experiments: More details}\label{sec:gen_models}

To model the piece-wise linear SoC charging curves, we break the charging into 3 segments. The charging power is kept constant at maximum (20 kW) for SoC $\leq$83\%, then linearly decreasing as $-\frac{4}{3} \cdot \text{SoC} + 130$ for 83–90\%, and $-\text{SoC} + 100$ for 90–100\%. Discharging assumes constant power with linear SoC decline. Electricity prices follow Silicon Valley Power rates: \$0.178/kWh peak (6 AM–10 PM), \$0.137/kWh off-peak, and an \$11.67/kW demand charge \cite{svp}.

\subsection{Forecast Generation. }\label{subsec:forecast_gen} 
For each control step, the MC-MPC framework requires forecasts of building load and EV behavior (arrivals, departures, SoC requirements).  
Building load forecasts are generated using standard autoregressive models fitted to the past month of building-meter data; electricity prices follow pre-published power utility rates and are assumed to be known a day ahead.

\noindent{\bf EV Behavior Forecast. }
Figure~\ref{fig:car_arrival_departure} presents the ground truth data for electric vehicle (EV) arrivals and departures, while Figure~\ref{fig:soc} illustrates the distributions of State of Charge (SoC) at both arrival and departure. EV behavior is modeled using generative probabilistic models trained on twenty-two months of real-world telemetry data.
EV behavior is modeled with generative probabilistic models trained on twenty-two months of telemetry data. Arrival counts per 15-minute interval are predicted using a zero-inflated negative binomial (ZINB) model, with input features including date, day of week, hour, and cyclical encodings of time-of-day (sine and cosine of hour).
Arrival SoC is modeled via a beta regression using day of week and cyclical time features.  
Departure SoC is predicted using a second beta regression conditioned on the arrival SoC, day of week and time features. 
Stay durations are drawn from a Weibull survival model with event rates dependent on day of week and arrival time.
We confirmed with the building operator that EV arrival and departure times depend solely on user schedules, not on building load or prior charging, so we presample arrivals, initial SoC, and dwell durations. This decoupling means charging decisions will not affect future availability, letting the controller focus exclusively on meeting SoC targets at minimum cost.

\paragraph{Model diagnostics.}
Table~\ref{tab:gen_metrics} lists concise goodness‑of‑fit statistics for each generative component that feeds the Monte‑Carlo SMPC.

\begin{table}[htbp]
\centering
\caption{Diagnostic metrics for generative EV behavior models}
\label{tab:gen_metrics}
\setlength\extrarowheight{2pt}
\small
\begin{tabular}{lccc}
\toprule
\textbf{Model} & \textbf{LL} & \textbf{AIC} & \textbf{Note} \\
\midrule
Arrival count (ZINB)    & $-2526$ & $5057$  & Pseudo-$R^2 = 0.34$ \\
Arrival SoC (Beta)      & $+128$  & $-241$  & Matches histogram well \\
Departure SoC (Beta)    & $+909$  & $-1801$ & Cond. on arrival improves fit \\
Stay duration (Weibull) & $-104$ & $-201$ & C-index $= 0.71$ \\
\bottomrule
\end{tabular}
\end{table}

\begin{itemize}
\item \textbf{Arrival counts.} The zero‑inflated negative‑binomial model explains roughly one‑third of the deviance (pseudo‑$R^{2}=0.34$), confirming that calendar and diurnal features capture meaningful variation in arrivals.  
\item \textbf{Arrival \& departure SoC.} Both beta regressions achieve strongly negative AIC values, indicating parsimonious yet well‑fitting density estimates for battery state distributions.
\item \textbf{Stay duration.} The Weibull survival model attains an AIC near –200 and a concordance index around 0.71, reflecting useful predictive skill for dwell times.  
\end{itemize}

These diagnostics support the credibility of the sampled future scenarios used by the MC-MPC controller.

\noindent{\bf Building Load Forecast.} 
Building demand can usually be predicted to within {$\approx$2–10\%} mean absolute percentage error (MAPE) over day‑ahead horizons\cite{cai2019day,yildiz2017review}.
We therefore create a “forecast’’ by adding zero‑mean Gaussian noise, tuned to that error band, to the realized load.
This keeps EV arrivals and departures as the dominant source of uncertainty and avoids entangling them with exogenous load‑forecast errors.
A fully coupled forecasting–control stack is left for future work.

\section{Results: More Discussion}
\begin{table*}[htbp]
\caption{Comparative analysis of different real-time charging policies, with costs scaled for consistency with Table~\ref{tab:combined_results}. The results highlight the superiority of the \textbf{MC-MPC} strategy, which achieves the lowest average electricity cost among all practical methods, performing within a small margin of the theoretical Oracle (MILP) benchmark. Crucially, \textbf{MC-MPC} accomplishes this economic efficiency without compromising service quality, delivering perfect charging reliability (zero unfulfilled SoC). It also provides the most effective peak shaving, demonstrating a robust balance of cost savings, reliability, and grid stability.}
\label{tab:combined_metrics_scaled}
\centering
\setlength\extrarowheight{2pt} %
\resizebox{\textwidth}{!}{
\begin{tabular}{p{2cm}|cccccccc|c}
\toprule
\textbf{Policy} & \textbf{May} & \textbf{Jun} & \textbf{Jul} & \textbf{Aug} & \textbf{Sep} & \textbf{Oct} & \textbf{Nov} & \textbf{Dec} & \textbf{8 months} \\
\midrule
\textit{Oracle (MILP)} & 7700 $\pm$ 118 & 8293 $\pm$ 101 & 8300 $\pm$ 114 & 12015 $\pm$ 87 & 8716 $\pm$ 68 & 9278 $\pm$ 92 & 8072 $\pm$ 111 & 6639 $\pm$ 110 & 8645 $\pm$ 1367 \\
\midrule
MC-MPC & \textbf{7776 $\pm$ 100} & {8516 $\pm$ 148} & {8622 $\pm$ 122} & {12121 $\pm$ 114} & \textbf{8828 $\pm$ 76} & \textbf{9384 $\pm$ 96} & \textbf{8169 $\pm$ 119} & \textbf{6862 $\pm$ 136} & \textbf{8785 $\pm$ 1429} \\
RL & {7817 $\pm$ 188} & \textbf{8446 $\pm$ 108} & \textbf{8512 $\pm$ 154} & \textbf{12052 $\pm$ 105} & {8896 $\pm$ 101} & {9398 $\pm$ 94} & {8221 $\pm$ 123} & {6899 $\pm$ 134} & {8808 $\pm$ 1441} \\
LLF & 7928 $\pm$ 146 & 8613 $\pm$ 135 & 8684 $\pm$ 208 & {12294 $\pm$ 92} & 8896 $\pm$ 118 & 9452 $\pm$ 114 & 8320 $\pm$ 134 & 7013 $\pm$ 142 & 8899 $\pm$ 1457 \\
EDF & 7933 $\pm$ 159 & 8609 $\pm$ 136 & 8687 $\pm$ 206 & 12294 $\pm$ 94 & 8898 $\pm$ 120 & 9453 $\pm$ 116 & 8321 $\pm$ 135 & 7014 $\pm$ 138 & 8900 $\pm$ 1457 \\
ReqCharge & 7945 $\pm$ 130 & 8667 $\pm$ 138 & 8731 $\pm$ 138 & 12428 $\pm$ 155 & 9009 $\pm$ 108 & 9548 $\pm$ 131 & 8385 $\pm$ 104 & 7085 $\pm$ 166 & 8975 $\pm$ 1482 \\
MaxCharge & 8345 $\pm$ 240 & 9002 $\pm$ 261 & 9112 $\pm$ 279 & 12496 $\pm$ 199 & 9313 $\pm$ 242 & 9615 $\pm$ 156 & 8670 $\pm$ 271 & 7506 $\pm$ 265 & 9257 $\pm$ 1389 \\
\bottomrule
\end{tabular}
}
\end{table*}

\begin{table*}[htbp]
  \caption{Base-case performance of Real‐Time Policies for all months}
  \label{tab:mpc_result_others}
  \scriptsize
  \centering
  \begin{minipage}{0.95\textwidth}
    \scriptsize
    *Higher peak shaving is better. Note that \textit{Oracle (MILP)} is only provided as a reference; it is an omniscient control strategy and is not practically feasible.
  \end{minipage}
  
  \vspace{1ex}
  \setlength{\tabcolsep}{3pt} %
  \setlength\extrarowheight{2pt}
  
  \begin{tabularx}{\textwidth}{l|*{7}{>{\centering\arraybackslash}X}}
    \toprule
    \textbf{Scenario} 
      & \textit{Oracle (MILP)} 
      & \textbf{MC-MPC} 
      & \textbf{RL} 
      & \textbf{LLF} 
      & \textbf{EDF} 
      & \textbf{ReqCharge} 
      & \textbf{MaxCharge} \\
    \midrule
    Missing SoC/month (in kWh)
      & 0.0 $\pm$ 0.0 
      & \textbf{0.00 $\pm$ 0.00} 
      & \textbf{0.00 $\pm$ 0.00} 
      & 3.66 $\pm$ 6.04 
      & 3.64 $\pm$ 5.81 
      & \textbf{0.00 $\pm$ 0.00} 
      & \textbf{0.00 $\pm$ 0.00} \\
  
    Cars with unfulfilled SoC
      & 0.0 $\pm$ 0.0 
      & \textbf{0.00 $\pm$ 0.00} 
      & \textbf{0.00 $\pm$ 0.00} 
      & 0.70 $\pm$ 0.97 
      & 0.69 $\pm$ 0.93 
      & \textbf{0.00 $\pm$ 0.00} 
      & \textbf{0.00 $\pm$ 0.00} \\
   
    Peak Shaving* (kW)
      & 28.37 $\pm$ 17.11 
      & \textbf{27.91 $\pm$ 16.82} 
      & 27.42 $\pm$ 16.42 
      & 15.04 $\pm$ 16.15 
      & 14.95 $\pm$ 16.14 
      & 9.25 $\pm$ 17.02 
      & N/A \\
    \bottomrule
  \end{tabularx}
\end{table*}

We begin the discussion by presenting a comprehensive evaluation of our proposed MC-MPC charging framework. This includes detailed performance metrics across key dimensions such as building cost, missing SoC, and peak shaving. By highlighting both aggregate trends and scenario-specific behaviors, we aim to demonstrate the robustness and adaptability of our approach under varying operational conditions.

\subsection{MC-MPC as Charging Policy}

\noindent{\bf Baseline Approaches.} We evaluate the performance of our online approach by comparing it against various methods including real-world charging procedures, several smart heuristic approaches, and a reinforcement learning-based policy. We provide a brief description of the baselines here.

\begin{itemize}
    \item {\bf MaxCharge}: This approach simulates current real-world charging, where all connected EVs are charged at the fastest possible rate.
    \item {\bf ReqCharge}: Similar to \textit{MaxCharge}, however, this policy only charges all connected EVs as quickly as possible to their requested SoC, $e^v_{req}$.
    \item {\bf Least Laxity First (LLF)}: A heuristic that prioritizes charging EVs with the least amount of idle time before their departure~\cite{nakahira2017smoothed}.
    \item {\bf Early Deadline First (EDF)}: A heuristic that prioritizes charging EVs with the earliest departure time~\cite{stankovic_EDF}.
    \item{\bf Reinforcement Learning (RL)}: We use a Deep Deterministic Policy Gradient (DDPG) algorithm to manage charging actions, as detailed in prior work~\cite{liu2025}. A separate model is trained for each month.
\end{itemize} 

We evaluate all approaches on $50$ test episodes per month on four key metrics: \emph{Total Cost}, \emph{Missing SoC}, the count of \emph{Cars Under Required SoC}, and \emph{Peak Shaving} relative to the MaxCharge policy.

\begin{table*}[htbp]
  \centering
  \caption{Electricity Cost (USD) of all Policies under Various Uncertainty Scenarios.}
  \label{tab:robustness_summary_elec}
  \scriptsize
  \vspace{1ex}
  \setlength{\tabcolsep}{2pt} %
  \setlength\extrarowheight{2pt}
  
  \begin{tabularx}{\textwidth}{l|*{7}{>{\centering\arraybackslash}X}}
    \toprule
    \textbf{Scenario} 
      & {\textit{Oracle (MILP)}} 
      & \textbf{MC-MPC} 
      & \textbf{RL} 
      & \textbf{LLF} 
      & \textbf{EDF} 
      & \textbf{ReqCharge} 
      & \textbf{MaxCharge} \\
    \midrule
    Departure-time uncertainty 
      & 7086 $\pm$ 1411 
      & \textbf{7334 $\pm$ 1480} 
      & 7357 $\pm$ 1445 
      & 7428 $\pm$ 1423 
      & 7433 $\pm$ 1420 
      & 7534 $\pm$ 1477 
      & 8059 $\pm$ 1409 \\
      
    OOD departure SoC 
      & 6882 $\pm$ 1401 
      & \textbf{7073 $\pm$ 1503} 
      & 7101 $\pm$ 1453 
      & 7203 $\pm$ 1464 
      & 7203 $\pm$ 1463 
      & 7246 $\pm$ 1514 
      & 7414 $\pm$ 1467 \\
      
    Building load noise 
      & 8281 $\pm$ 1489 
      & 8534 $\pm$ 1627 
      & \textbf{8519 $\pm$ 1529} 
      & 8605 $\pm$ 1584 
      & 8606 $\pm$ 1586 
      & 8699 $\pm$ 1639 
      & 8964 $\pm$ 1585 \\
    \midrule
    Mid-day price spike 
      & 7040 $\pm$ 1378 
      & \textbf{7238 $\pm$ 1521} 
      & 7248 $\pm$ 1499 
      & 7348 $\pm$ 1501 
      & 7349 $\pm$ 1500 
      & 7395 $\pm$ 1560 
      & 7580 $\pm$ 1506 \\
      
    Fleet scale-up 
      & 7089 $\pm$ 1487 
      & \textbf{7288 $\pm$ 1560} 
      & N/A 
      & 7348 $\pm$ 1501 
      & 7349 $\pm$ 1500 
      & 7395 $\pm$ 1560 
      & 7580 $\pm$ 1506 \\
    \bottomrule
  \end{tabularx}
\end{table*}

\begin{table*}[htbp]
  \centering
  \caption{Cumulative Missing SoC per month (kWh) over all Policies under Various Uncertainty Scenarios}
  \label{tab:robustness_summary_missing_soc}
  \scriptsize
  \setlength\extrarowheight{2pt}
  \begin{tabularx}{\textwidth}{l|*{7}{>{\centering\arraybackslash}X}}
    \toprule
    \textbf{Scenario} 
      & {\textit{Oracle (MILP)}} 
      & \textbf{MPC} 
      & \textbf{RL} 
      & \textbf{LLF} 
      & \textbf{EDF} 
      & \textbf{ReqCharge} 
      & \textbf{MaxCharge} \\
    \midrule

    Departure‐time uncertainty
      & {9.31 $\pm$ 23.09} 
      & \textbf{9.31 $\pm$ 23.09}
      & {10.29 $\pm$ 23.09}
      & 15.12 $\pm$ 23.55
      & 15.43 $\pm$ 23.55
      & \textbf{9.31 $\pm$ 23.09}
      & \textbf{9.31 $\pm$ 23.09} \\
    OOD departure SoC
       & {0.49 $\pm$ 1.07} 
      & \textbf{0.49 $\pm$ 1.07}
      & 9.45 $\pm$ 5,44
      & 3.95 $\pm$ 6.08
      & 3.85 $\pm$ 5.91
      & \textbf{0.49 $\pm$ 1.07}
      & \textbf{0.49 $\pm$ 1.07} \\
    Building load noise
      & 0.00 $\pm$ 0.00
      & 26.50 $\pm$ 18.45
      & 26.50 $\pm$ 18.45
      & {2.94 $\pm$ 5.09}
      & 2.96 $\pm$ 5.13
      & \textbf{0.00 $\pm$ 0.00}
      & \textbf{0.00 $\pm$ 0.00}  \\

  \midrule
  
    Mid-day price spike
      & 7040 $\pm$ 1378
      & \textbf{7238 $\pm$ 1521}
      & 7248 $\pm$ 1499
      & 7348 $\pm$ 1501
      & 7349 $\pm$ 1500
      & 7395 $\pm$ 1560
      & 7580 $\pm$ 1506 \\

    Fleet scale-up
       & 0.00 $\pm$ 0.00
      & \textbf{0.00 $\pm$ 0.00}
      & N/A
      & 2.94 $\pm$ 5.13
      & 2.96 $\pm$ 5.16
      & \textbf{0.00 $\pm$ 0.00}
      & \textbf{0.00 $\pm$ 0.00} \\
    
    \bottomrule
  \end{tabularx}
  \\[1ex]
\begin{minipage}{0.95\textwidth}
  \footnotesize
   Note that \textit{Oracle (MILP)} is only provided as a reference; it is an omniscient control strategy that is not practically feasible.
\end{minipage}
\end{table*}

\noindent{\bf Comparative Performance Analysis.}
Table~\ref{tab:combined_metrics_scaled} reports the scaled monthly electricity costs for each policy. The offline {Oracle (MILP)}, while infeasible in practice, provides a theoretical lower bound on performance. Our {MC-MPC} policy performs exceptionally well, with an 8-month average cost of \$8785, which is within {1.6\%} of the oracle's benchmark. This demonstrates a near-optimal balance of foresight and real-time adaptability.

The {RL} policy achieves a competitive average cost of \$8808. While it secures lower costs in some summer months by implicitly learning temporal demand patterns, our {MC-MPC} proves more stable and cost-effective on average. Both intelligent policies significantly outperform the heuristics; {MC-MPC} is approximately {1.3\%} more efficient than {LLF}~\cite{nakahira2017smoothed} and {EDF}~\cite{stankovic_EDF}. The gap is even wider when compared to rigid, real-world policies, with our approach outperforming {ReqCharge} and {MaxCharge} by {2.2\%} and {5.4\%}, respectively.

Beyond monthly costs, reliability and grid impact are critical. Through Table~\ref{tab:mpc_result_others}, it is  observed that {MC-MPC} and {RL} consistently meet all EV energy demands, resulting in zero missing SoC across all test episodes. In contrast, heuristic approaches like LLF and EDF fail to fully charge several vehicles each month. Furthermore, {MC-MPC} achieves the highest peak shaving among all practical methods, outperforming RL slightly and surpassing the heuristics by a wide margin. These metrics confirm that {MC-MPC} offers the most robust and reliable solution, delivering near-optimal cost savings without compromising user satisfaction or grid stability.

\noindent{\bf Experiments Under Uncertainty. } \label{subsec:uncertainty} 
Uncertainty is a fundamental aspect of real-world EV charging operations. Vehicle arrivals and departures are often unpredictable, user preferences can vary dynamically, and building loads fluctuate throughout the day. To evaluate the robustness of different control strategies under these practical conditions, we assess performance across five representative uncertainty scenarios. These scenarios encompass both aleatoric uncertainty, arising from inherent randomness, and epistemic uncertainty, stemming from model misalignment or deployment shifts. Tables~\ref{tab:robustness_summary_elec} and~\ref{tab:robustness_summary_missing_soc} summarize each strategy’s ability to handle uncertainty, measured in terms of total building cost and the incidence of missing state-of-charge (SoC) values.

\noindent{\bf Aleatoric uncertainty:} 
    \begin{itemize}
    \item \textbf{Departure-time uncertainty}: EV departure deviation modeled as a Gaussian perturbation with mean as the initial departure time, and 2-hour standard deviation applied to departure times, given they have enough time to charge to meet the required SoC.
  \item \textbf{Out-of-distribution (OOD) SoC requests}: Departure SoC targets are perturbed by adding a random offset, uniformly sampled within $\pm$15\% of the originally generated SoC value, simulating deviations from the assumed generative model.

    \item \textbf{Building load noise}: The building’s baseline demand is perturbed by sampling zero-mean noise from a uniform distribution with support $\pm$ 15\% of the average daily load. This models day-level fluctuations in building load without adding a systematic bias.
    \end{itemize}

Across all scenarios in Table~\ref{tab:robustness_summary_elec}, our proposed MC-MPC controller exhibits consistent and stable performance, incurring only 2–3\% additional electricity cost compared to the offline MILP oracle. This reflects its ability to plan over sampled futures and adjust to emerging disturbances, without requiring retraining or explicit reconfiguration. 
While the reinforcement learning (RL) approach demonstrates strong performance under some scenarios, particularly under building load noise, they show higher variance and less reliable SoC fulfillment in others, such as out-of-distribution SoC requests. This brittleness under distribution shift is expected: RL policies are typically trained on a fixed data distribution and can overfit to nominal dynamics, making them less robust in deployment. 
Heuristic methods (EDF, LLF) and fixed rules (ReqCharge, MaxCharge) suffer in both cost and reliability. They lack foresight and adaptability, resulting in higher cumulative shortfalls in energy delivery and frequent under-charging of vehicles, especially when operating outside their nominal range.

\noindent{\bf Epistemic and structural uncertainty:} 
    \begin{itemize}
    \item \textbf{Mid-day price spike}: Electricity tariffs are artificially increased by 20\% during the mid-day peak period (12 PM to 4 PM), simulating a sudden price surge.
    \item \textbf{Fleet scale-up}: The number of chargers and simultaneously connected EVs is increased by a factor of 2, stressing the system’s scalability and responsiveness.
    \end{itemize} 
Table~\ref{tab:robustness_summary_missing_soc} further illustrates MC-MPC’s advantage: even under temporal or preference variability, it meets SoC targets without compromise. This is achieved by jointly optimizing shared first-stage decisions across multiple plausible futures, rather than reacting myopically or committing prematurely. Such lookahead, combined with real-time feedback and constraint-awareness, enables MC-MPC to generalize across operational scenarios.

Overall, these results highlight the strength of MC-MPC as a general-purpose, real-time decision-making framework: it scales to large fleets, adapts to non-stationarity, and remains resilient under uncertainty, offering a strong alternative to brittle learning methods and inflexible heuristics in V2B coordination.

\begin{table}[htbp]
\centering
\small
\caption{MC-MPC Hyperparameter Search Results}
\begin{tabular}{lccccc}
    \toprule
    \textbf{Param} & \textbf{Description} & \textbf{Min} & \textbf{Max} & \textbf{Step} & \textbf{Best} \\
    \midrule
    $\mathcal{F}$ & Sample size & 5 & 50 & 5 & 30 \\
    $P_{DC}$ & Demand penalty & 0 & 50 & 5 & 10 \\
    $P_{SOC}$ & SoC penalty & 0 & 20 & 1 & 1 \\
    $q$ & Max rate change & 0 & 10 & 2.5 & 5 \\
    \bottomrule
\end{tabular}
\label{tab:hyperparameter_optimization_performance}
\end{table}

\subsection{User Choices and Negotiation Behavior}

We aim to understand how simulated users interact with the menu of charging options in our negotiation framework, through Figures~\ref{fig:choice_dist},~\ref{fig:dep_dist}, and~\ref{fig:soc_dist}.

Figure~\ref{fig:choice_dist} indicates that most users gravitate toward the \textit{exact} fulfillment option (Option 0), suggesting a strong preference for maintaining their original charging request without deviation. Nonetheless, while the majority remain inflexible, a substantial subset is willing to accept one of the flexible options (Choices 1, 2, or 3) or even to reject all offers. This highlights the dual reality faced by system designers: exact fulfillment cannot be relied upon exclusively for system flexibility, yet there remains meaningful engagement with negotiated alternatives, providing valuable operational leeway for building operators.

\noindent\textbf{Operational Deviations: Departure and SoC}

Figure~\ref{fig:dep_dist} shows the distribution of departure time deviations across all users. The sharp spike at zero deviation confirms that most users are not shifted from their requested departure, with only small fractions accepting nontrivial delays. Notably, the presence of a secondary group experiencing larger delays underscores that operational flexibility is selectively leveraged: the system offers minimal inconvenience to most, while extracting valuable participation from a subset of more flexible users. The wide tail further emphasizes the heterogeneity in user flexibility and tolerated inconvenience.

\begin{figure*}[htbp]
    \centering
    \begin{subfigure}[b]{0.32\textwidth}
        \centering
        \includegraphics[width=\linewidth]{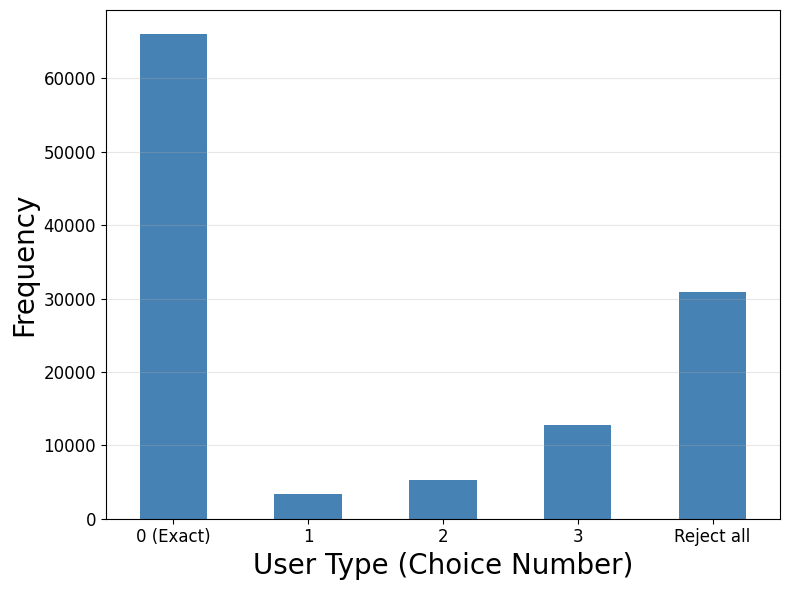}
        \caption{Choice distribution}
        \label{fig:choice_dist}
    \end{subfigure}
    \hfill
    \begin{subfigure}[b]{0.32\textwidth}
        \centering
        \includegraphics[width=\linewidth]{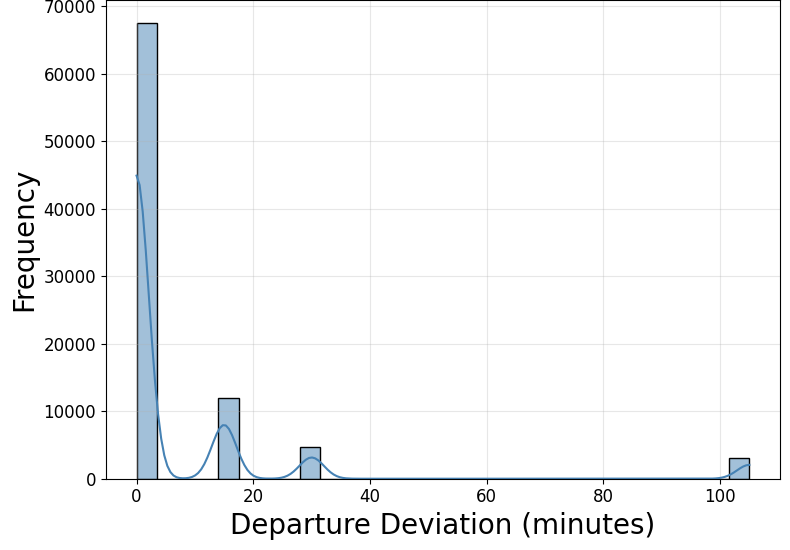}
        \caption{Departure deviation}
        \label{fig:dep_dist}
    \end{subfigure}
    \hfill
    \begin{subfigure}[b]{0.32\textwidth}
        \centering
        \includegraphics[width=\linewidth]{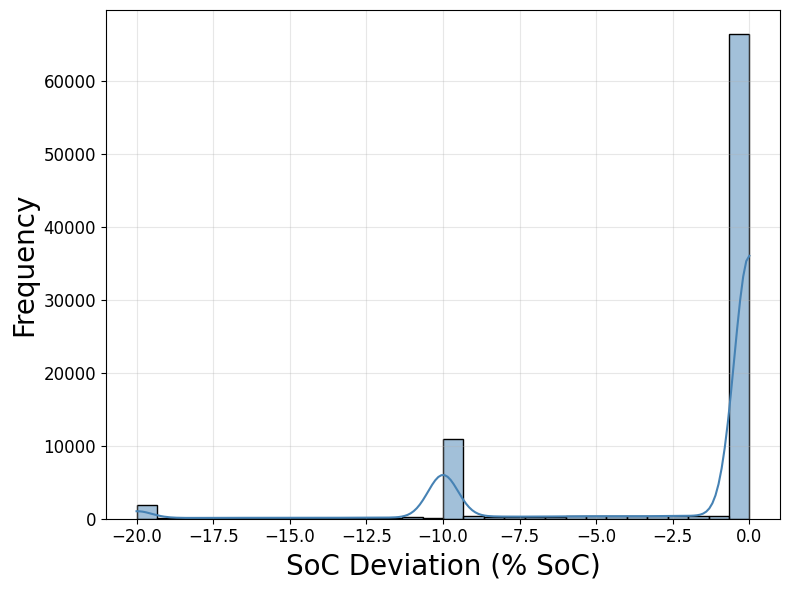}
        \caption{SoC deviation}
        \label{fig:soc_dist}
    \end{subfigure}
    
    \caption{Histograms of user negotiation behavior: choice distribution, departure time deviation, and SoC deviation for all users.}
    \label{fig:combined_distributions}
\end{figure*}

\begin{figure*}[htbp]
    \centering
    \includegraphics[width=0.85\textwidth, height=5.5cm]{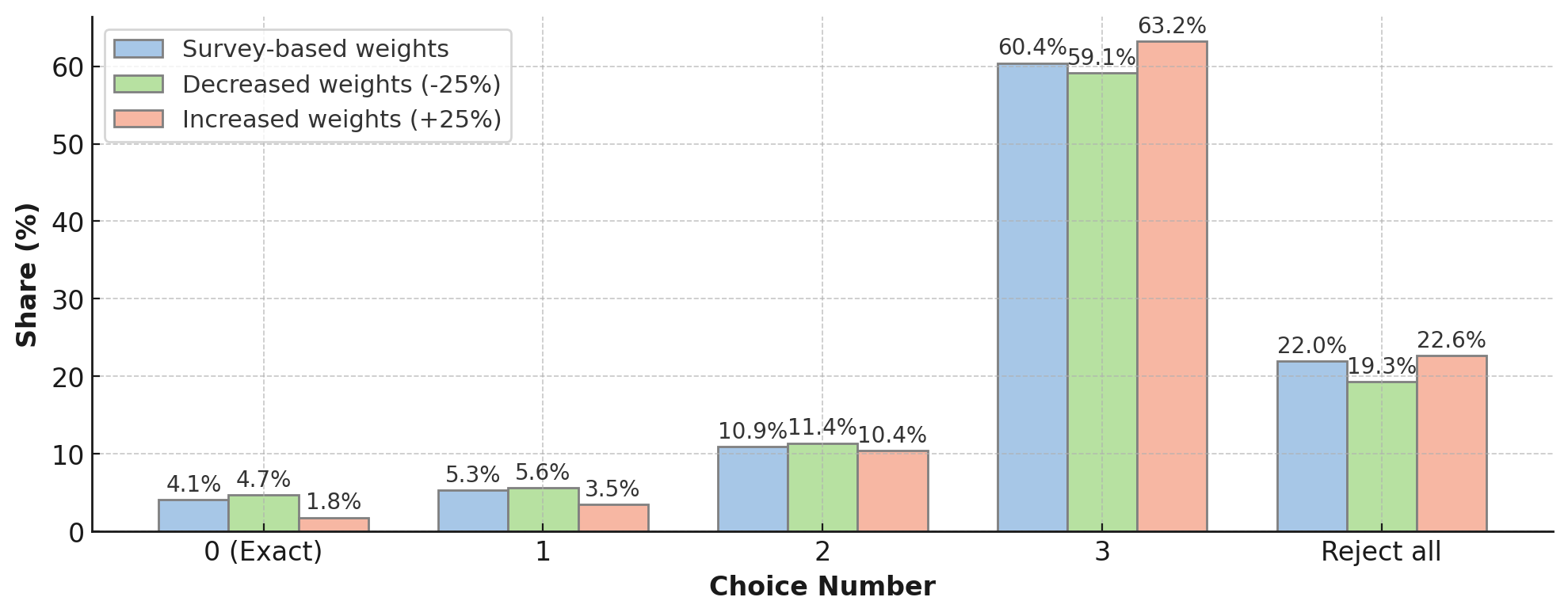}
    \caption{User cost and distribution upon changing  the user sensitivity weights using CONSENT}
    \label{fig:change_user_type}
\end{figure*}

Similarly, Figure~\ref{fig:soc_dist} shows the distribution of SoC deviations at departure. Note that these deviations are represented as negative values relative to the user's requested SoC, and thus the data appear primarily in the second quadrant of the plot.
Nearly all users experience negligible SoC shortfalls, shown by the pronounced peak near zero deviation. However, isolated secondary peaks at moderate and larger SoC deficits reveal that some negotiated solutions do require a minority of users to accept tangible compromises. The majority's satisfaction is preserved, while building-level optimization gains cost-effective flexibility by dispersing minor inconveniences among the most adaptable users.

\noindent\textbf{Analysis of Users Choosing to ``Reject''}
Figure~\ref{fig:reject-cost_utility} presents the distribution of user cost and user satisfaction \(Y\) (as defined in eq.~\eqref{eq:user_satisfaction}) for each negotiation option, including those users who choose to reject all offers. The left figure shows that user costs remain relatively consistent across options \(l=0\) through \(l=3\), with only moderate variation. Crucially, even for users selecting the ``Reject All'' option, the average simulated user cost is still notably lower than the external charging benchmark cost.

This outcome is statistically robust due to the large number of Monte Carlo samples used in the simulation, ensuring convergence of empirical averages to expected values as per the law of large numbers. Consequently, the distribution of costs observed accurately reflects the underlying probabilistic model of user behavior and system dynamics, validating that participation in the negotiation process offers economic benefits on average, even when users ultimately reject all negotiated offers.

In the right figure, user satisfaction reveals a more nuanced pattern. Users who choose either the exact solution ($l=0$) or the ``Reject All'' option tend to report the lowest satisfaction scores, reflecting limited perceived value or compensation for accommodating flexibility. However, in several cases, higher flexibility options (such as $l=3$) could have delivered slightly higher user satisfaction, suggesting that some users may not be fully accounting for the additional utility available through more flexible solutions.

This result underscores two important points for negotiation-based charging strategies. First, user cost savings alone are not sufficient to guarantee satisfaction and participation, especially for those with limited willingness to defer or compromise. Second, improved communication or interface design could help users recognize and capitalize on more rewarding (but flexible) options, ultimately improving their experience and system participation rates.

Ultimately, these findings highlight the importance of aligning incentive structures and user communication, ensuring that users not only realize economic savings, but also perceive adequate compensation for their flexibility, especially among more discriminating or conservative user segments.

\begin{figure*}[htbp]
    \centering
    \includegraphics[width=\textwidth]{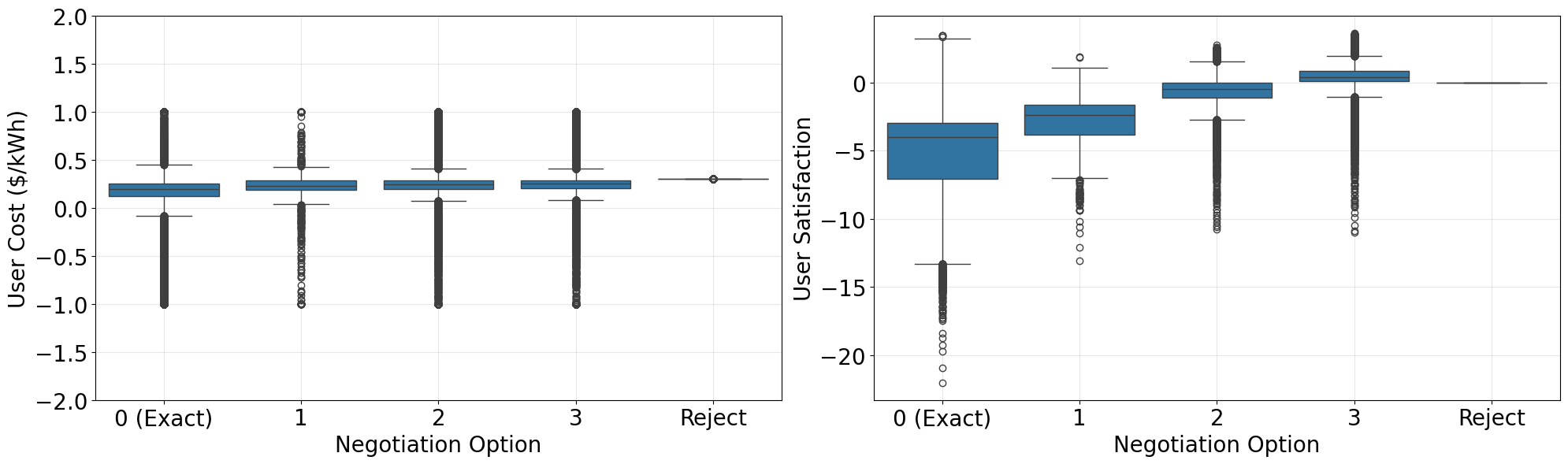}
    \caption{User cost and utility distribution by negotiation option (for users who chose to reject)}
    \label{fig:reject-cost_utility}
\end{figure*}

Complementing these metrics, Figure~\ref{fig:change_user_type} illustrates the distribution of user choices across negotiation options under the three scenarios. Key observations include:

\begin{itemize}
    \item \textbf{Exact Choice (Choice 0):} The share of users selecting their exact requested profile decreases notably when sensitivity increases (from 4.1\% to 1.8\%), while it slightly increases under reduced sensitivity (to 4.7\%). This indicates that higher sensitivity discourages strict adherence to original preferences.
    
    \item \textbf{Flexible Deviations (Choices 1–3):} The uptake of flexible negotiated options fluctuates moderately, with the most flexible option (Choice 3) gaining share as sensitivity increases (from 59.1\% to 63.2\%) and losing some share under reduced sensitivity (to 59.1\%), suggesting that even sensitive users may accept greater deviations in specific cases.
    
    \item \textbf{Rejection (Choice 5):} Rejection rates rise from 20.5\% at baseline to 21.0\% with increased sensitivity and drop to 17.9\% with reduced sensitivity, reinforcing that users who perceive greater inconvenience are more likely to reject offers.
\end{itemize}

In summary, these findings demonstrate that user education or interventions that effectively reduce perceived sensitivity to flexibility can lower system costs and increase participation, while increases in sensitivity generally have the opposite effects. The observed shifts in choice distribution emphasize the behavioral mechanisms influencing negotiation outcomes. These trade-offs should be carefully considered when designing demand-side management programs to optimize both system efficiency and user satisfaction.

\noindent \textbf{More Details on Sensitivity to Cost-Sharing Levels.}  
Table~\ref{tab:cost_sharing} evaluates our negotiation framework under varying proportions of cost-sharing between the building and users, from full (100\%) sharing down to minimal (10\%) sharing of savings. As the building’s share decreases, we observe a steady decrease in building costs, reaching a minimum reduction of approximately 6.5\% at 10\% sharing, representing higher direct building savings.

Conversely, user costs increase with lower sharing percentages, indicating a shift of cost burden to users. User satisfaction and system utilization also decline gradually as sharing decreases, reflecting reduced incentives and acceptance under higher individual costs. 

This sensitivity analysis highlights the trade-offs inherent in cost allocation within the negotiation design: higher building savings come at the expense of user costs and engagement. The selection of an appropriate sharing level thus balances operational savings with user participation and satisfaction, providing flexibility to adapt to deployment priorities.

Overall, these results complement our core findings by demonstrating the negotiation mechanism’s robustness and tunability across diverse cost-sharing preferences.

\noindent\textbf{Interpretation and System-Level Tradeoffs}

Collectively, these empirical distributions underscore several important findings:
\begin{itemize}
    \item \textbf{User Heterogeneity}: Most users are relatively inflexible, but a significant flexible minority provides enough operational slack for meaningful negotiation outcomes.
    \item \textbf{Limited Impact on Most Users}: The negotiation framework maintains negligible inconvenience for the majority, ensuring high overall acceptance and satisfaction.
    \item \textbf{Minority Flexibility Drives Savings}: Substantial cost savings and peak-shaving benefits are made possible by the minority willing to accept time or SoC deviations. The system strategically allocates inconvenience with minimal overall dissatisfaction.
    \item \textbf{Participation–Cost Tradeoff}: As further quantified in sensitivity tables and rejection analysis, user participation and satisfaction are maximized when incentives are substantial and deviation boundaries are carefully set; overly aggressive cost-shifting suppresses engagement.
\end{itemize}

In summary, the distributions make plain that our negotiation-based framework achieves a robust balance: it delivers operational savings comparable to more coercive or top-down methods, but maintains high user acceptance and satisfaction by leveraging heterogeneous flexibility and tailoring minor concessions primarily to those most willing or able to absorb them. These results complement the core findings on cost-sharing and demonstrate the framework’s adaptability for real-world deployment.

\section{Stated-preference Survey}\label{sec:survey_full}

\subsection{Section A}

Thank you for choosing to take this survey. Your perspective is valuable - even if you don't
currently drive an EV. To help, we'll provide a set of simple assumptions for you to base your 
responses on. Please answer using your best understanding and estimates.
We're developing an EV charging system that lets you earn discounts by being a little flexible,
either by departing a bit later or accepting a slightly lower battery level. Although you can
charge at home or at public stations, this survey focuses on how workplace and school
charging can reward that flexibility. Your feedback will help us create a more sustainable,
grid-friendly charging experience for everyone. To give you an idea of how this system works,
let's walk through an example.
Imagine arriving at work/school each morning, parking in your usual spot, and plugging in
your EV. On a EV companion app, you enter the driving range (or battery charge level) you
need when you leave and your planned departure time. Right before charging begins, the app
presents three tailored offers on your phone. A few possible choices:

\begin{itemize}
    \item Leave on time with a full charge - regular price, say \$10.
    \item Leave a bit later, or take slightly less range - think of it like filling up a bit less fuel on a gas car in exchange for a moderate discount, say \$8 (The discount is greater than the amount of charge/fuel you give up).
    \item Leave up to an hour later, or accept a more noticeable reduction in range - like topping off your gas tank with less fuel but paying much less, say \$6 (The discount is much greater than the amount of charge/fuel you give up).
    \item If you think the price is too high, you can reject and not choose to charge at school/work. The nice thing is, you always have the option to come back next time and try a different choice.

\end{itemize}

When you pick one of the offers, you're making a single-day "deal" for that charging session.

\vspace{5mm}

\textbf{1. How do you typically commute to your workplace or school?}
\begin{itemize}
    \item I drive (my own car or a shared one)
    \item Public transportation (bus, train, tram, etc.)
    \item Bicycle/Bike
    \item Walk
    \item Motorcycle/Scooter
    \item Rideshare (e.g., Uber, Lyft)
    \item I primarily work/study remotely and do not have a regular commute
    \item Other
\end{itemize}

\textbf{2. What is your experience with electric vehicles (EVs)?}
\begin{itemize}
    \item I currently own, regularly drive and charge an EV.
    \item I have driven and charged an EV one or more times.
    \item I have an understanding of how EV charging works and have been a passenger in one.
    \item I have not had any personal experience driving or charging an EV.
    \item I am not sure.
\end{itemize}

\textbf{3. What time do you typically arrive at your work or school on a regular day?}\\
(Time hh:mm input)
\vspace{8pt}

\textbf{4. And, how many hours would you typically spend at work or school on such a day?}\\
(Numerical input)
\vspace{8pt}

\textbf{5. How many miles do you usually travel to and from work or school?}\\
(Numerical input)
\vspace{8pt}

\subsection{Section B}

For the following questions, you can base your answers on these Scenario Assumptions:

You own an EV, and a full battery gives about 300 miles of range (like a full gas tank).
Work/school provides EV charging stations for employee/student use. You have a charger
installed at home (used for cost comparisons). Charging prices get lower the more flexibility
(departure time and EV driving range) you accept. Your anonymous answers in this survey
tell us how much flexibility and compensation feel fair, so we can provide the user with fair
compensation for their flexibility and design a user-friendly system. Please remember that
your honest feedback in answering all the questions is valuable, whether positive or negative;
we're not testing you, and we welcome all insights to help us improve.

\textbf{6. On most days, what's the minimum EV range (in miles) you need before your next charging session at work/school?}
\begin{itemize}
    \item Less than 20 miles (very short direct commute home)
    \item 20 to 39 miles
    \item 40 to 59 miles
    \item 60 to 79 miles
    \item 80 to 99 miles
    \item 100 to 119 miles
    \item More than 120 miles
    \item I find it difficult to estimate this.
\end{itemize}

\textbf{7. On a typical day, what is the maximum delay you would comfortably accept to reduce your charging costs?}
\begin{itemize}
    \item 0 minutes, leave exactly on time
    \item 15 minutes
    \item 30 minutes
    \item 45 minutes
    \item 60 minutes
    \item 75 minutes
    \item 90 minutes
\end{itemize}

\textbf{8. Driving a 300-mile range EV, imagine you arrive with 30\% charge and want your car charged to 80\% by departure time. Which discount options would you consider fair if you left with less charge?}
\begin{itemize}
    \item I insist on my ideal 80\% (240 miles range) charge.
    \item Accept leaving with $\sim$75\% charge (approx.\ 225 miles range, so 15 miles less) for a small benefit.
    \item Accept leaving with $\sim$70\% charge (approx.\ 210 miles range, so 30 miles less) for a fair benefit.
    \item Accept leaving with $\sim$60\% charge (approx.\ 180 miles range) for a significant benefit.
    \item I need more information to answer.
\end{itemize}

\textbf{9. Please specify the ideal cost reduction (in dollars) you would expect for the benefit level you chose in the previous question, Q8 (small, fair, or significant benefit):} \\
(If you are answering using a mobile device, tap on the slider square to activate it, and then move it) \\
\vspace{8pt}

\textbf{10. As a quick check to confirm you're following along, please select option B below.} \\
\begin{itemize}
    \item A
    \item B
    \item C
    \item D
\end{itemize}

\subsection{Section C}

In this section, we ask you to imagine owning an EV with a 300-mile battery. You arrive with a 30\% charge (about 90 miles) and aim to reach 80\% (240 miles) by the time you leave. A full charge in this scenario normally costs \$10.

You will now be asked:

How much compensation would you consider fair if the system asked you to either delay your departure time or accept a slightly lower charge?
This is part of a smart negotiation system: The higher the compensation you request, the less likely it is that the system accepts your
offer. In some cases, no compensation may be offered if the system determines that your
requested compensation is too high.

Note: The \$10 cost already covers battery degradation and maintenance. Any discount you select will be applied in addition to this amount. This representative \$10 cost is typically lower than public charging rates and may also be more cost-effective than home charging in certain markets.
Your answers below help us understand what people consider fair, enabling us to design
systems that work better for both users and the grid.

\textbf{11. 30 minutes delay: If you were asked to delay your departure from work/school by 30 minutes and a full charging session normally costs \$10.00, which of these discounts would feel fair?} \\
\begin{itemize}
    \item No monetary incentive needed - I would do it to support sustainability efforts.
    \item 5\% off - Almost always accepted
    \item 10\% off - Very likely to be accepted
    \item 20\% off - Likely to be accepted
    \item 25\% off - May be accepted in some cases
    \item 30\% off - Unlikely to be accepted
    \item 35\% off - Very unlikely to be accepted
    \item 40\% off or more - Extremely unlikely to be accepted
    \item I would not delay my departure for 30 minutes for an incentive of this scale.
    \item Not applicable / Cannot estimate.
\end{itemize}

\textbf{12. 15 minutes delay: If you were asked to delay your departure from work/school by 15 minutes and a full charging session normally costs \$10.00, which of these discounts would feel fair?} \\
\begin{itemize}
    \item No monetary incentive needed - I would do it to support sustainability efforts.
    \item 5\% off - Almost always accepted
    \item 10\% off - Very likely to be accepted
    \item 20\% off - Likely to be accepted
    \item 25\% off - May be accepted in some cases
    \item 30\% off - Unlikely to be accepted
    \item 35\% off - Very unlikely to be accepted
    \item 40\% off or more - Extremely unlikely to be accepted
    \item I would not delay my departure for 15 minutes for an incentive of this scale.
    \item Not applicable / Cannot estimate.
\end{itemize}

\textbf{13. 60 minutes delay: If you were asked to delay your departure from work/school by 60 minutes, and a full charging session normally costs \$10.00, which of these discounts would feel fair?} \\
\begin{itemize}
    \item No monetary incentive needed - I would do it to support sustainability efforts.
    \item 5\% off - Almost always accepted
    \item 10\% off - Very likely to be accepted
    \item 20\% off - Likely to be accepted
    \item 25\% off - May be accepted in some cases
    \item 30\% off - Unlikely to be accepted
    \item 35\% off - Very unlikely to be accepted
    \item 40\% off or more - Extremely unlikely to be accepted
    \item I would not delay my departure for 60 minutes for an incentive of this scale.
    \item Not applicable / Cannot estimate.
\end{itemize}

\textbf{14. 70\% instead of 80\%: Imagine your EV has a 300-mile battery. You plug in with 30\% charge (90 miles) and want to charge to 80\% (240 miles) for \$10. Today, the system offers to charge you instead to 70\% (210 miles) for \$8. You would still have enough range to get home and run errands. What is the smallest extra discount off this \$8 charge that would feel fair for the reduced range?} \\
\begin{itemize}
    \item No monetary incentive needed - if the reduced range is sufficient.
    \item 5\% off - Very likely to be accepted
    \item 10\% off - Likely to be accepted
    \item 15\% off - May be accepted in some cases
    \item 20\% off - Unlikely to be accepted
    \item 25\% off - Very unlikely to be accepted
    \item 30\% off or more - Extremely unlikely to be accepted
    \item I would not accept less range for an incentive of this scale.
    \item Not applicable / Cannot estimate.
\end{itemize}

\textbf{15. 75\% instead of 80\%: Similar to Q14, you plug in with 30\% charge and want 80\% charge and the charging cost is shown as \$10. Today, the system provides you an alternative and offers to charge you instead to 75\% (225 miles) for \$9. You would still have enough range to get home and run errands. What is the smallest extra discount off this \$9 charge that would feel fair for the reduced range?} \\
\begin{itemize}
    \item No monetary incentive needed - if the reduced range is sufficient.
    \item 5\% off - Very likely to be accepted
    \item 10\% off - Likely to be accepted
    \item 15\% off - May be accepted in some cases
    \item 20\% off - Unlikely to be accepted
    \item 25\% off - Very unlikely to be accepted
    \item 30\% off or more - Extremely unlikely to be accepted
    \item I would not accept less range for an incentive of this scale.
    \item Not applicable / Cannot estimate.
\end{itemize}

\textbf{16. 60\% instead of 80\%: Similar to Q14, you plug in with a 30\% charge and want an 80\% charge, and the charging cost is shown as \$10. Today, the system provides you another alternative to 60\% (180 miles) for \$6. You would still have enough range to get home and run errands. What is the smallest extra discount off this \$6 charge that would feel fair for the reduced range?} \\
\begin{itemize}
    \item No monetary incentive needed - if the reduced range is sufficient.
    \item 5\% off - Very likely to be accepted
    \item 10\% off - Likely to be accepted
    \item 15\% off - May be accepted in some cases
    \item 20\% off - Unlikely to be accepted
    \item 25\% off - Very unlikely to be accepted
    \item 30\% off or more - Extremely unlikely to be accepted
    \item I would not accept less range for an incentive of this scale.
    \item Not applicable / Cannot estimate.
\end{itemize}

\textbf{17. General Feedback:} \\

\end{document}